\documentclass[twocolumn]{aastex631}

\usepackage{amsmath}
\usepackage{amssymb} 
\usepackage{epstopdf}
\usepackage{graphicx}
\usepackage{float}
\usepackage{subfigure}
\usepackage{indentfirst}
\usepackage{color}
\usepackage{natbib}
\usepackage[figuresright]{rotating}
\usepackage{enumerate}
\newcommand{\continuedcaption}[1]{
  \addtocounter{figure}{-1} 
  \caption{(Continued) #1} 
}

\shortauthors{Zhang et al.}

\begin{document}

\title{Spatially-resolved Galactic HII regions observed by LAMOST medium-Resolution Spectroscopic Survey of Nebulae (LAMOST MRS-N)}


\author[0000-0002-1783-957X]{Wei Zhang$^\dagger$}
\affiliation{CAS Key Laboratory of Optical Astronomy, National Astronomical Observatories,  Chinese Academy of Sciences, Beijing 100101, People's Republic of China}
\email{Wei Zhang$^\dagger$: xtwfn@bao.ac.cn}

\author[0009-0000-6954-9825]{Yunning Zhao}
\affiliation{CAS Key Laboratory of Optical Astronomy, National Astronomical Observatories, Chinese Academy of Sciences, Beijing 100101, People's Republic of China}
\affiliation{School of Astronomy and Space Science, University of Chinese Academy of Sciences, Beijing 100049, China}

\author{Lin Ma}
\affiliation{{The Key Laboratory of Cosmic Rays (Tibet University), Ministry of Education, Lhasa 850000, Tibet, China}}
\affiliation{CAS Key Laboratory of Optical Astronomy, National Astronomical Observatories, Chinese Academy of Sciences, Beijing 100101, People's Republic of China}
\affiliation{School of Astronomy and Space Science, University of Chinese Academy of Sciences, Beijing 100049, China}

\author[0009-0008-1361-4825]{Shiming Wen} 
\affiliation{CAS Key Laboratory of Optical Astronomy, National Astronomical Observatories, Chinese Academy of Sciences, Beijing 100101, People's Republic of China}

\author{Chaojian Wu}
\affiliation{CAS Key Laboratory of Optical Astronomy, National Astronomical Observatories, Chinese Academy of Sciences, Beijing 100101, People's Republic of China}

\author{Juanjuan Ren}
\affiliation{CAS Key Laboratory of Space Astronomy and Technology, National Astronomical Observatories, Chinese Academy of Sciences, Beijing 100101, People's Republic of China}

\author{Jianjun Chen}
\affiliation{CAS Key Laboratory of Space Astronomy and Technology, National Astronomical Observatories, Chinese Academy of Sciences, Beijing 100101, People's Republic of China}

\author{Yuzhong Wu}
\affiliation{CAS Key Laboratory of Optical Astronomy, National Astronomical Observatories, Chinese Academy of Sciences, Beijing 100101, People's Republic of China}

\author{Zhongrui Bai}
\affiliation{CAS Key Laboratory of Optical Astronomy, National Astronomical Observatories, Chinese Academy of Sciences, Beijing 100101, People's Republic of China}

\author{Yonghui Hou}  
\affiliation{School of Astronomy and Space Science, University of Chinese Academy of Sciences, Beijing 100049, China}
\affiliation{Nanjing Institute of Astronomical Optics, \& Technology, National Astronomical Observatories, Chinese Academy of Sciences, Nanjing 210042, People's Republic of China}

\author{Yongheng Zhao}
\affiliation{CAS Key Laboratory of Optical Astronomy, National Astronomical Observatories, Chinese Academy of Sciences, Beijing 100101, People's Republic of China}
\affiliation{School of Astronomy and Space Science, University of Chinese Academy of Sciences, Beijing 100049, China}

\author{Hong Wu}
\affiliation{CAS Key Laboratory of Optical Astronomy, National Astronomical Observatories, Chinese Academy of Sciences, Beijing 100101, People's Republic of China}
\affiliation{School of Astronomy and Space Science, University of Chinese Academy of Sciences, Beijing 100049, China}

        
\label{firstpage}

\begin{abstract}
We present spatially-resolved spectroscopic observations of 10 isolated Galactic HII regions using data from the LAMOST Medium-Resolution Spectroscopic Survey of Nebulae (LAMOST MRS-N). The high spatial resolution of the data allows us to investigate the 1D radial profiles of emission line fluxes (H$\alpha$, [S II] and [N II]), flux ratios  ([N II]/H$\alpha$, [S II]/H$\alpha$ and [S II]/[N II]), and radial velocities of these three emission lines. Among these regions, two are ionization-bounded, while the remaining eight are matter-bounded. The matter-bounded HII regions exhibit shallower slopes in their radial flux profiles compared to the ionization-bounded ones. In most cases, the [N II]/H$\alpha$ and [S II]/H$\alpha$ ratios increase with distance from the center of the HII regions, consistent with model predictions that low-ionization emissions dominate the outer zones of these regions. The two ionization-bounded HII regions have kinematic ages $t$ of 0.2 and 0.3 Myr, while the matter-bounded regions span ages from 1 to 12 Myr. For the matter-bounded HII regions, the optical emission flux decreases continuously beyond the photodissociation region (PDR), extending to approximately 1-4 times the radius of the PDR (r$_{PDR}$). The escape fraction $f_{esc}$ of ionizing photons, derived from 1D H$\alpha$ radial flux profiles, is $\sim$ 0\% for ionization-bounded HII regions, while it ranges from 50\% to 90\% for the matter-bounded HII regions. The correlation between $f_{esc}$ and $t$ suggests that evolved HII regions  (with $t > $ 1 Myr) contribute more significantly to ionizing the surrounding diffuse ionized gas (DIG) compared to younger, newly formed HII regions.
\end{abstract}


\section{Introduction}

In addition to HII regions, diffuse ionized gas (DIG) represents a significant component of the interstellar medium in the Milky Way \citep{Reynolds-1984} and external spiral galaxies. Studies suggest that DIG contributes a substantial fraction, ranging from 30\% to 60\%, of the total H$\alpha$ emission \citep{Oey-2007}. While photons leaking from HII regions are considered the primary source of ionization, DIG is not confined to spiral galaxies. It has also been observed in early-type galaxies, which typically exhibit lower ongoing star formation rates \citep{Phillips-1986, Martel-2004, Jaffe-2014, Johansson-2016}. This indicates that the ionization and maintenance of DIG are not solely driven by active star formation but likely involve additional physical processes. These processes may include photoionization by older stellar populations (HOLMES) \citep{Binette-1994, Stasinska-1994, Athey_Bregman-2009, Flores-Fajardo-2011, Yan_Blanton-2013, Zhang-2017}, cosmic rays \citep{Reynolds_Cox-1992, Vandenbroucke-2018}, shocks \citep{Martin-1997, Collins_Rand-2001, Ramirez-Ballinas-2014}, and dust scattering \citep{Barnes-2015, Ascasibar-2016}.

An intriguing question arises concerning the evolving interaction between HII regions and the surrounding DIG. Using radio data, researchers have analyzed the ionization profiles of HII regions and made notable discoveries. Large HII regions, for instance, exhibit shallow ionization profiles \citep{Luisi-2019}, suggesting that a significant fraction of photons escape from the partially ionized photodissociation regions (PDRs), thereby injecting additional energy into the DIG. Furthermore, recent simulations indicate that very young HII regions (with ages less than 5 Myr) contribute minimally to the ionization of the DIG. Instead, the primary sources of ionizing photons are relatively older HII regions (ages between 5 and 25 Myr), which play a key role in clearing the surrounding interstellar medium (ISM) and sustaining the ionization of the DIG \citep{McClymont-2024}.

To investigate the evolution of ionization profiles and the changing dynamics of interactions, spatially-resolved 2D spectra of HII regions are essential. HII regions typically span scales of 1-100 pc, but current large-scale integral field unit (IFU) surveys, such as CALIFA \citep{Sanchez-2012} and MaNGA \citep{Bundy-2015}, operate at spatial resolutions on the order of kpc. This makes it challenging to achieve the necessary spatial resolution for detailed analysis of individual HII regions. In this work, we overcome this limitation by selecting a sample of Galactic HII regions from the Large Sky Area Multi-Object Fiber Spectroscopic Telescope (LAMOST) dataset \citep{Wang-96, Su-Cui-04, Cui-12, Zhao-12, Luo-15}. These regions are characterized by large apertures and high spatial resolution, enabling us to explore the intricate details of ionization profiles and kinematic properties across different evolutionary stages. We note that, based on the LAMOST dataset, \cite{Wen-2025} compiled a large sample of DIG in the anticenter region of the Milky Way and investigated the radial and vertical distributions of three line ratios ([N II]/H$\alpha$, [S II]/H$\alpha$, and [S II]/[N II]) as well as the oxygen abundance. They found that [N II]/H$\alpha$ and [S II]/H$\alpha$ exhibit enhancements within the interarm region, located between the Local Arm and the Perseus Arm. Additionally, the oxygen abundance displays a consistent radial gradient with galactocentric distance, with no evidence of flattening in the outer disk.

The article is structured as follows. Section 2 provides a brief introduction to the data used and outlines the criteria for sample selection. Section 3 details the data analysis, including measurements of distance, physical size, extinction, emission line flux and radial velocity, ionization and velocity profiles, as well as calculations of the escape fraction, expansion velocity, and kinematic age. The results and discussion are presented in Section 4, and the conclusions are summarized in the final section.

\begin{figure*}[htbp]
\centering
\includegraphics[width=\textwidth]{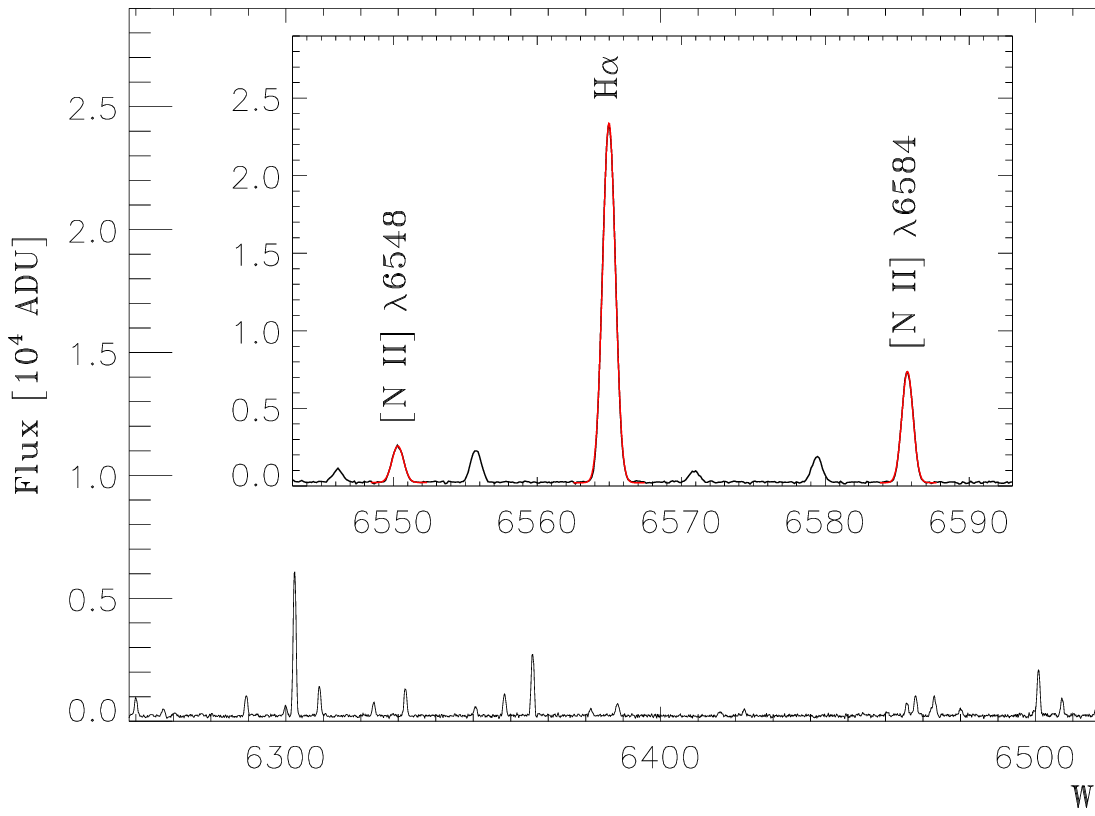}
\caption{A representative spectrum observed by LAMOST MRS-N. Nebular emission lines, including H$\alpha$, [N II], and [S II], are enlarged in the inset panels. The results of single Gaussian fitting are overplotted in red. As the spectrum is not flux-calibrated, the flux is given in units of Analog-to-Digital Units (ADU).
\label{fig:example_spectrum}}
\end{figure*}

\section{sample construction}

Based on mid-infrared (MIR) morphology observed by the all-sky Wide-Field Infrared Survey Explorer (WISE) satellite \citep{Wright-2010}, \cite{Anderson-2014} compiled the largest catalog of Galactic HII regions. We use Version 2.2 of this catalog, which is publicly available on their website (http://astro.phys.wvu.edu/wise/). Out of 8412 cataloged sources, 2210 have been classified as ``Known HII regions" based on detections in radio recombination lines (RRLs) or H$\alpha$ emission. For each HII region, Anderson's catalog provides the circular radius that encloses the associated MIR emission. In our analysis, we adopt this radius as the boundary radius (r$_{PDR}$) of the observed HII regions.

LAMOST, also known as the Guo Shou Jing Telescope, is a 4-meter quasi-meridian reflective Schmidt telescope located at the Xinglong Observatory in China. With a field of view (FoV) of 5$^{\arcdeg}$, LAMOST is equipped with 4000 fibers on its focal plane, enabling simultaneous spectral observations of nearly 4000 celestial objects. This unique combination of a large aperture and wide FoV makes LAMOST a powerful facility for large-scale spectral surveys \citep{Wang-96, Su-Cui-04, Cui-12, Zhao-12, Luo-15}. LAMOST operates in two resolution modes: low-resolution (R $\sim$ 1800) and medium-resolution (R $\sim$ 7500). The Low-Resolution Spectroscopic Survey (LRS) covers a spectral wavelength range of 3700–9000 Å, while the Medium-Resolution Spectroscopic Survey (MRS) spans 4950–5350 \AA (blue segment) and 6300–6800 \AA (red segment) \citep{Liu-2020}.

The MRS was launched in October 2018 with the goal of achieving precise radial velocity (RV) measurements ($\rm \sim 1~km\,s^{-1}$ for late-type stars) and exploring a wide range of scientific topics, including binarity, stellar pulsation, star formation, emission nebulae, Galactic archaeology, exoplanet host stars, and open clusters \citep{Liu-2020}. \cite{ZhangB-2021} measured the RVs of 3.8 million single-exposure MRS stellar spectra (for 0.6 million stars) with signal-to-noise ratios (S/N) greater than 5 from LAMOST DR7 using the cross-correlation function method. They found that the precision of the RVs is approximately 0.38 $\rm km\,s^{-1}$ for RV zero-points (RVZPs)  achieved with the help of Gaia DR2 RVs. Validated with APOGEE DR16 data, the absolute RVs achieve an overall precision ranging from 0.80 to 1.99 $\rm km\,s^{-1}$, depending on the S/N of the spectra. For 678 standard stars with multiple observations, the standard deviations of the RVs range from 0.54 to 1.86 $\rm km\,s^{-1}$ across different confidence levels.

As part of the MRS, the LAMOST Medium-Resolution Survey of Galactic Nebulae (LAMOST MRS-N) specifically targets emission nebulae within the Galactic plane, covering a longitude range of 40\arcdeg\, $<$ l $<$ 215\arcdeg\, and $|b| < $ 5\arcdeg, corresponding to a sky coverage of $\sim$ 1700 deg$^2$. The data products include several optical emission lines, such as H$\alpha$, [N II], [S II], and [O III] \citep{Wu-2021}.  The wavelengths were re-calibrated using sky lines, which revealed systematic RV deviations ranging from $\sim$ 0.2 to 0.5 $\rm km\,s^{-1}$ by the end of 2018 \citep{Ren-2021}. Geocoronal H$\alpha$ emission (H$_{\alpha,sky}$) has been subtracted by using the correlation between the line ratio of H$_{\alpha,sky}$ to the OH $\lambda$6544 skyline and solar altitude \citep{ZhangW-2021}. The data processing pipeline and products are summarized in \cite{Wu-2022}.

Each fiber in the LAMOST focal plane has an aperture size of 3$\farcs$3 in diameter. The spatial resolution is determined by the median fiber-to-fiber separation in a single exposure, which is $\sim$ 3$\arcmin$ \citep{Wen-2025}. In overlapping regions, the spatial resolution is higher, and in some specific areas with multiple observations, it can reach about 0$\farcm$5 \citep{Wu-2021}.

For this study, we focus exclusively on the red segment of the spectrum. A representative spectrum is shown in Figure \ref{fig:example_spectrum}, where the H$\alpha$, [N II], and [S II] emission lines are overplotted with the results of Gaussian fitting, highlighted in red. For clarity, these nebular lines are enlarged in the inset panels.

We performed a cross-match between the LAMOST MRS-N data and the WISE HII region catalog, applying the following selection criteria:
\begin{itemize}
\item [1)] The region must be classified as a ``Known HII region".
\item [2)] No bright HII regions should be located within twice the radius of r$_{PDR}$.
\item [3)] The angular size of the region must be sufficiently large ($r > $ 3\farcm5) to ensure that at least several fibers are assigned within the circular region defined by r$_{PDR}$.
\end{itemize}
After applying these criteria, we identified and selected 10 ``Known HII regions".  The corresponding WISE catalog IDs, coordinates of the HII region centers, and r$_{PDR}$ values (in arcminutes) are listed in Table \ref{tbl:pars}.

\begin{figure*}[htbp]
\centering
\includegraphics[width=\textwidth]{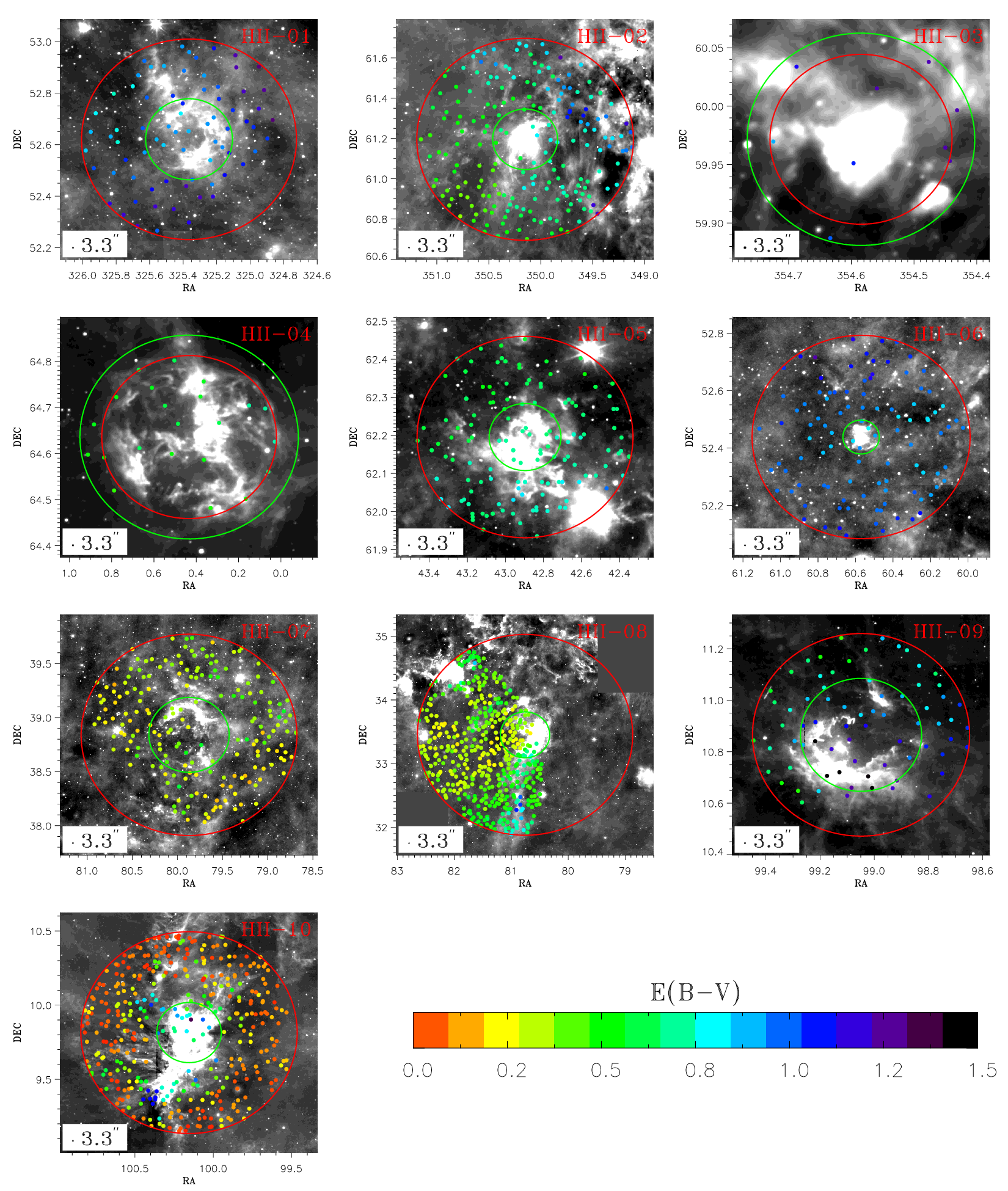}
\caption{Positions of the LAMOST fibers marked with dotes, color coded based on reddening E(B-V) derived from the Galactic 3D reddening map by \cite{Green-2019}. For clarity, the sizes of these fibers are enlarged, while the actual angular size (3$\farcs$3 in
diameter) is shown as black dot in the bottom-left corner. Green circles represent sizes of HII regions ($r_{PDR}$) defined by the WISE catalog, while red circles indicate radii ($r_{H\alpha}$) derived from 1D H$\alpha$ intensity profiles in this work. Background images are from WISE 12 $\mu$m observations.}
\label{fig:wise_img}
\end{figure*}

The final sample is illustrated in Figure \ref{fig:wise_img}, where the background image represents the WISE 12 $\mu$m emission, and the dots indicate the positions of the LAMOST MRS-N fibers. The actual angular size of the fibers (3$\farcs$3 in diameter) is shown as a black dot in the bottom-left corner, color-coded by reddening E(B-V). The green circles depict the radii of the HII regions (r$_{PDR}$), as defined by \cite{Anderson-2014} using WISE images. The red circles represent the regions influenced by the HII regions, as determined by the 1D radial profile of H$\alpha$ fluxes (see Section \ref{sec:flux_profile} for detailed explanations).

\section{data analysis}

\subsection{Distance and physical size}

According to the WISE HII catalog, only one HII region in our sample has a parallax distance measurement, while seven have kinematic distances derived from the RV of RRLs or H$\alpha$ emission lines. For the remaining two HII regions, neither RV nor distance information is provided in the WISE HII catalog. Although LAMOST MRS-N also provides RV measurements of optical emission lines, which can be used to calculate kinematic distances, the uncertainties associated with this method are significant, reaching up to approximately 50\% in the anti-center direction \citep{Anderson-2014}. To achieve more accurate distance estimates, we instead use parallax distances to massive OB stars or young stellar objects (YSOs) associated with the HII regions.

In this study, we first identify OB stars or young stellar objects (YSOs) located within one $r_{PDR}$ radius for each HII region. The OB stars are selected from the OB star catalog \citep{Liu-2019,Xu-2021} or retrieved from the SIMBAD database, while YSOs are obtained exclusively from the SIMBAD database. \cite{Bailer-Jones-2021} derived two types of parallax distances for stars in Gaia Early Data Release 3: geometric distances, which use a direction-dependent prior, and photogeometric distances, which additionally incorporate stellar color and apparent magnitude. Photogeometric distances generally provide better accuracy and precision, especially for stars with less reliable parallax measurements. Their catalog includes 1.35 billion photogeometric distances, along with asymmetric uncertainty estimates. We cross-match the identified OB stars and YSOs with this catalog to determine the distances to the HII regions. If only one OB star or YSO is found within an HII region, we adopt its parallax distance as the distance to the corresponding HII region. For regions with multiple OB stars or YSOs, we calculate the angular distance ($\theta$) to the center of the HII region and use $\theta^{-2}$ as a weighting factor to compute the weighted distance, which is then assigned as the distance to the HII region. Using this approach, we successfully determined distances for all ten HII regions in our sample. The most distant HII region is located at 5.3 kpc, while the nearest one lies at a distance of 0.7 kpc from the observer.

Furthermore, we estimate the physical sizes of the HII regions by combining their distances with the angular sizes provided by WISE. The results are listed in Table \ref{tbl:pars}. The physical sizes (2r$_{PDR}$) range from 4.8 pc to 32.6 pc. Notably, the typical size of HII regions is approximately 30 pc \citep{Weaver-1977}.

\subsection{Extinction}

The ionized gas in HII regions is often associated with molecular gas \citep{Anderson-2009}, and foreground molecular gas can also affect observations of emission lines. Since molecular gas is not expected to be uniformly distributed, it is necessary to estimate the extinction for each fiber individually. The most accurate method for this is the Balmer decrement method. However, H$\beta$ falls outside the wavelength coverage of the LAMOST MRS-N, making this approach unfeasible. Instead, we rely on the Galactic reddening map. A 2D reddening map is unsuitable for this purpose, as these HII regions are located in the Galactic disk, and a 2D map provides only the total extinction along each line of sight. Therefore, we use a 3D reddening map \citep{Green-2019} to account for the spatial distribution of extinction.

For simplicity, we adopt a single distance, as calculated earlier, to derive the E(B-V) values, which represent the total reddening between the HII regions and the observer. The 2D distribution of these E(B-V) values is shown in Figure \ref{fig:wise_img}. This distribution reveals diverse extinction patterns. For some HII regions, the extinction is low and relatively uniform, while for others, it is higher and exhibits more pronounced spatial variations. These variations further underscore the necessity of applying extinction corrections.

\subsection{Integrated flux and radial velocity}

For each nebular emission line, we fit it with a single Gaussian profile to simultaneously determine the line centroid, line dispersion, and integrated flux. The fitting results for these lines in a representative spectrum are shown in red in Figure \ref{fig:example_spectrum}. Currently, the LAMOST MRS-N data have not been flux-calibrated, and the integrated flux of these lines is in units of Analog-to-Digital Units (ADU).

We apply the extinction law from \cite{Fitzpatrick-1999}, adopting the default value of R$_V$ = 3.1. In the following text, we define  F(H$\alpha$), F([N II]), and F([S II]) as the dereddened fluxes of the H$\alpha$, [N II] $\lambda$6584, and [S II] $\lambda$6717 emission lines, respectively. Additionally, we define N2Ha, S2Ha, and S2N2 as the flux ratios F([N II])/F(H$\alpha$), F([S II])/F(H$\alpha$), and F([S II])/F([N II]), respectively.

RVs are derived from the difference between the line centroid and the rest-frame wavelength of each emission line. The local standard of rest (LSR) velocity (V$_{LSR}$)is then calculated by correcting the RV for the motion of the Sun relative to the LSR. The Sun is moving at a speed of approximately 20 $\rm km\,s^{-1}$ toward the direction (RA = 18$^h$03$^m$50$^s$29, DEC = +30\arcdeg00\arcmin16$\farcs$8) at epoch 2000.

\begin{figure*}
\centering
\includegraphics[width=\textwidth]{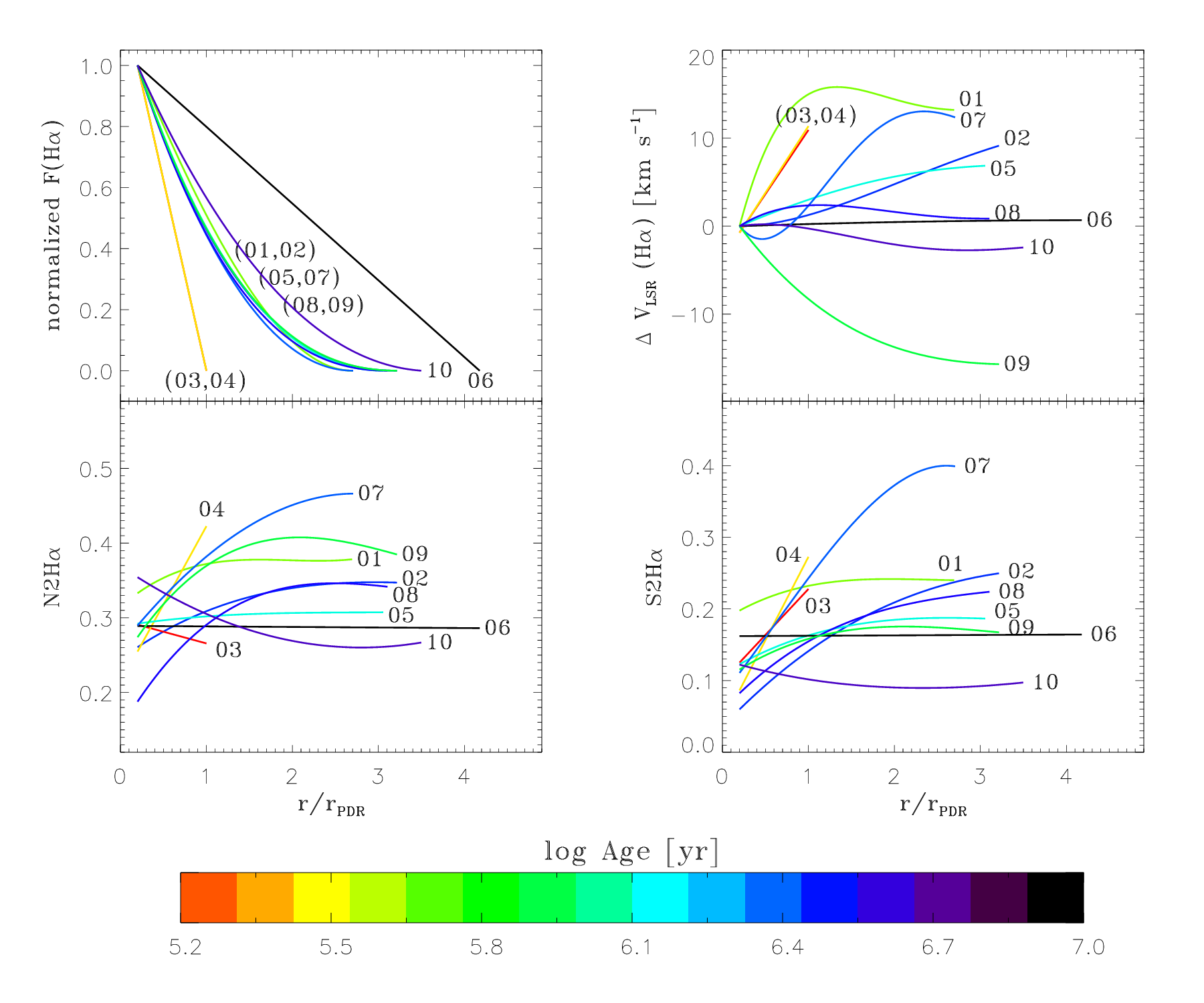}
\caption{Radial profiles of the normalized H$\alpha$ flux, relative H$\alpha$ radial velocity, and flux ratios, color-coded by kinematic age. These curves are derived from the fitting results shown in Figures \ref{fig:flux_radius}, \ref{fig:flux_ratio_radius}, and \ref{fig:vlsr_radius}. In the upper-left panel, the H$\alpha$ flux is normalized to its value at $ r = 0.2r_{PDR}$. In the upper-right panel, the V$_{LSR}$  of H$\alpha$ represents the relative velocity compared to the V$_{LSR}$  at $ r = 0.2r_{PDR}$. The ID number for each HII region is displayed alongside its corresponding curve. For curves that are difficult to distinguish, their ID numbers are grouped together and shown in parentheses.}
\label{fig:profiles}
\end{figure*}

\subsection{Ionization and velocity profiles}
\label{sec:profile}

The specific functional form of the flux and velocity of ionized gas as a function of distance from the center of an HII region (hereafter denoted as $r$) is crucial for understanding the underlying physical processes during the evolution of HII regions. 

The ionization profile can, in principle, be derived from narrow-band H$\alpha$ images; however, this approach faces challenges due to contamination from the [N II] doublet lines within the filter. Given the non-uniformity of the [N II]/H$\alpha$ ratio across HII regions, the H$\alpha$ profile cannot be accurately determined using narrow-band imaging alone. To address this limitation, 2D spectral observations allow for the separate extraction of the spatial distributions of H$\alpha$, [N II], and [S II] lines.

Using spectroscopic data instead of narrow-band images offers several advantages. First, it reduces uncertainties associated with subtracting background continuum components, a task more effectively accomplished with spectroscopic data. Second, it enables the examination of line ratio profiles as a function of distance from the HII region center. Third, spectroscopic data also provide insights into velocity profiles.

LAMOST MRS-N presents an excellent opportunity to comprehensively characterize ionization and velocity profiles in both 2D and 1D, offering detailed insights for a sample of HII regions with diverse physical sizes.

\subsubsection{Radial profile of line flux}
\label{sec:flux_profile}

In Figure \ref{fig:flux_radius}, we present the radial profiles of the integrated flux of three emission lines—H$\alpha$, [N II], and [S II]—as a function of $r$. It is evident that, for each HII region, the intensity of these emission lines gradually decreases with increasing distance from the center of the HII region. The data points, represented by crosses, correspond to measurements obtained from individual fibers. The typical 1$\sigma$ observational uncertainties are shown in the upper-right corner. For each HII region, we divided the data into 10 bins along the x-axis. Since the range (in units of arcminutes) varies among these HII regions, the bin size is not uniform across them. For each bin, we calculated the median value (represented by magenta dots) and the 1$\sigma$ scatter (magenta error bars), without considering the uncertainties of the individual data points.

Our analysis reveals that these profiles cannot be accurately described by power-law or exponential functions. Instead, we employ a polynomial function to fit the binned data. A default polynomial order of 4 was used for fitting. However, for HII-03, HII-04, and HII-06, the order was reduced to 2, as we found that the default order of 4 led to overfitting for these three targets. The detailed coefficients and corresponding errors are listed in Table \ref{tbl:coeff1}.

The red vertical lines indicate the critical radius, denoted as $r_{H\alpha}$, which marks the boundary at which the H$\alpha$ flux decreases the background value. The fundamental principle for determining r$_{H\alpha}$ is that at this radius, the H$\alpha$ flux decreases to the level of the background emission near each HII region. For the majority of the HII regions (8 out of 10), r$_{H\alpha}$ is determined by the turning point of a 4-order polynomial fit to the data. For  HII-03 and HII-04, where the H$\alpha$ flux declines sharply, a linear fit suffices. For these two regions, we visually selected the radial range of the data to be used for fitting. The values of r$_{H\alpha}$ can be found in Table \ref{tbl:coeff1}. For comparative analysis, we also show the radius $r_{PDR}$ from the WISE HII catalog as green vertical lines.

The radial profile of H$\alpha$ flux is then normalized to its value at 0.2$r_{PDR}$ for each HII region. All normalized profiles are shown in the upper-left panel of Figure \ref{fig:profiles} for comparison, with the x-axis converted to the normalized radius ($r/r_{PDR}$). The ID number of each HII region is displayed next to its corresponding curve. For curves that are difficult to distinguish, their ID numbers are grouped together and presented in parentheses. Among the ten HII regions studied, eight exhibit $r_{H\alpha}$ values exceeding $r_{PDR}$, indicating that a fraction of the ionizing energy leaks out of these regions. We classify these eight regions as matter-bounded HII regions. In contrast, the remaining two HII regions have $r_{H\alpha}$ values smaller than $r_{PDR}$, suggesting that most of the ionizing energy within the optical spectrum is confined within these regions. We therefore classify these two as ionization-bounded HII regions.

\begin{deluxetable*}{cccccccccccccccc}
\tablecaption{Summary of parameters for the ten HII regions. Col. 1 lists the IDs assigned to these regions, labeled from HII-01 to HII-10. Col. 2 provides the WISE names cataloged by \cite{Anderson-2014}. Cols. 3 to 6 present the radius in arcminutes, V$_{LSR}$ in $\rm km\,s^{-1}$, distance in kpc, and the method used to determine the distance, as sourced from \cite{Anderson-2014}. Col. 7 lists the V$_{LSR}$ of the H$\alpha$ line in the central region, while Col. 8 shows the V$_{LSR}$ at the PDR. Col. 9 details the expansion velocity of the HII shell, and Col. 10 provides distances derived from the central OB star using Gaia parallax. Col. 11 displays the physical radius as observed by WISE. Col. 12 represents the kinematic age, and Col. 13 indicates the escape fraction of ionizing photons. Col. 14 shows the optical aperture radius derived from the 1D H$\alpha$ flux profile, while Col. 15 denotes the optical physical radius of the H$\alpha$ emission. \label{tbl:pars}}
\tabletypesize{\tiny}
\tablehead{       
 \multicolumn{1}{c}{}  \vline & \multicolumn{5}{c}{WISE} \vline& \multicolumn{9}{c}{LAMOST MRS-N}  \\
\hline      
(1) & (2) & (3) & (4) & (5) & (6) & (7) & (8) & (9) & (10) & (11) &  (12) & (13) & (14) & (15)\\
\colhead{ID}          & 
WISE Name & \colhead{$r_{PDR}$} & \colhead{$V_{LSR}$}   & \colhead{dist}  &  \colhead{method} & 
\colhead{$V_{cen}$} & \colhead{$V_{PDR}$} & \colhead{$V_{exp}$} & \colhead{dist} & \colhead{$r_{PDR}$} & \colhead{Age} & \colhead{$f_{esc}$} & \colhead{$r(H\alpha$)} & \colhead{r(H$\alpha$)}\\
                        & 
                        & \colhead{[arcmin]} & \colhead{[$\rm km\,s^{-1}$]} & \colhead{[kpc]} &        & 
\colhead{[$\rm km\,s^{-1}$]}  & \colhead{[$\rm km\,s^{-1}$]}  & \colhead{[$\rm km\,s^{-1}$]}   & \colhead{[kpc]}  &  \colhead{[pc]} & \colhead{[Myr]} & & \colhead{[arcmin]} & \colhead{[pc]}
}                 
\startdata
  HII-01 & G096.345-00.157 &    9.4 &  -47.9 &       6.3 & OG/Kin     &  -34.2 &  -19.3 &   15.0 &    5.2 &   14.3 &    0.6 &  0.55 &   25.4 &   38.6 \\
  &   & - & - & $\pm$1.4 &   &  $\pm$1.2 &  $\pm$4.9 & $\pm$5.0 & $\pm$0.2 & $\pm$0.7 & $\pm$0.2 & $\pm$0.58 & - & - \\
  HII-02 & G112.212+00.229 &    9.0 &  -43.9 &       4.6 & OG/Kin     &  -45.1 &  -43.8 &    1.3 &    2.7 &    7.1 &    3.3 &  0.58 &   29.0 &   22.8 \\
  &   & - & - & $\pm$1.3 &   &  $\pm$0.1 &  $\pm$2.5 & $\pm$2.5 & $\pm$0.1 & $\pm$0.3 & $\pm$6.5 & $\pm$0.00 & - & - \\
  HII-03 & G113.900-01.613 &    5.4 &  -46.1 &       4.8 & OG/Kin     &  -43.8 &  -29.2 &   14.6 &    2.7 &    4.3 &    0.2 &  0.00 &    4.4 &    3.4 \\
  &   & - & - & $\pm$1.3 &   &  $\pm$3.9 &  $\pm$8.5 & $\pm$9.4 & $\pm$0.1 & $\pm$0.1 & $\pm$0.1 & - & - & - \\
  HII-04 & G117.639+02.275 &   13.3 &  -56.6 &       5.8 & OG/Kin     &  -53.9 &  -38.7 &   15.1 &    2.3 &    8.9 &    0.3 &  0.00 &   10.6 &    7.1 \\
  &   & - & - & $\pm$1.1 &   &  $\pm$2.4 &  $\pm$8.6 & $\pm$8.9 & $\pm$0.1 & $\pm$0.5 & $\pm$0.2 & - & - & - \\
  HII-05 & G136.448+02.519 &    5.3 &  -46.7 &       5.1 & OG/Kin     &  -45.6 &  -42.6 &    3.0 &    5.3 &    8.2 &    1.6 &  0.57 &   16.1 &   25.1 \\
  &   & - & - & $\pm$1.2 &   &  $\pm$0.4 &  $\pm$4.6 & $\pm$4.6 & $\pm$0.7 & $\pm$1.1 & $\pm$2.5 & $\pm$0.12 & - & - \\
  HII-06 & G149.738-00.207 &    3.5 &  -71.4 &       17.5 & OG/Kin     &  -18.6 &  -18.3 &    0.2 &    4.7 &    4.9 &   12.0 &  0.86 &   14.8 &   20.3 \\
  &   & - & - & $\pm$5.5 &   &  $\pm$1.2 &  $\pm$2.2 & $\pm$2.5 & $\pm$0.5 & $\pm$0.5 & $\pm$129.8 & $\pm$0.01 & - & - \\
  HII-07 & G168.750+00.873 &   20.8 &  -26.3 &            &            &  -21.6 &  -19.3 &    2.3 &    1.9 &   11.2 &    2.8 &  0.52 &   56.3 &   30.4 \\
  &   & - & - &   &   &  $\pm$4.9 &  $\pm$5.5 & $\pm$7.3 & $\pm$0.1 & $\pm$0.3 & $\pm$8.9 & $\pm$0.34 & - & - \\
  HII-08 & G173.588-01.606 &   21.6 &   -1.4 &            &            &   -6.1 &   -3.8 &    2.4 &    2.6 &   16.3 &    4.0 &  0.56 &   67.2 &   50.6 \\
  &   & - & - &   &   &  $\pm$1.0 &  $\pm$2.5 & $\pm$2.7 & $\pm$0.1 & $\pm$0.5 & $\pm$4.6 & $\pm$0.04 & - & - \\
  HII-09 & G201.535+01.597 &   13.2 &   23.1 &       3.9 & OG/Kin     &   17.7 &    9.4 &    8.3 &    3.6 &   13.7 &    1.0 &  0.58 &   42.3 &   43.9 \\
  &   & - & - & $\pm$1.7 &   &  $\pm$6.3 &  $\pm$5.5 & $\pm$8.3 & $\pm$0.1 & $\pm$0.6 & $\pm$1.0 & $\pm$0.07 & - & - \\
  HII-10 & G202.968+02.083 &   12.2 &    0.9 &       0.7 & Parallax   &   -0.6 &   -0.9 &    0.3 &    0.7 &    2.4 &    5.0 &  0.68 &   42.6 &    8.4 \\
  &   & - & - & $\pm$0.1 &   &  $\pm$9.7 &  $\pm$2.5 & $\pm$10.0 & $\pm$0.0 & $\pm$0.0 & $\pm$181.2 & $\pm$0.26 & - & - \\
\hline              
\enddata
\end{deluxetable*}  

\begin{figure*}[ht]
\centering
\includegraphics[width=\textwidth]{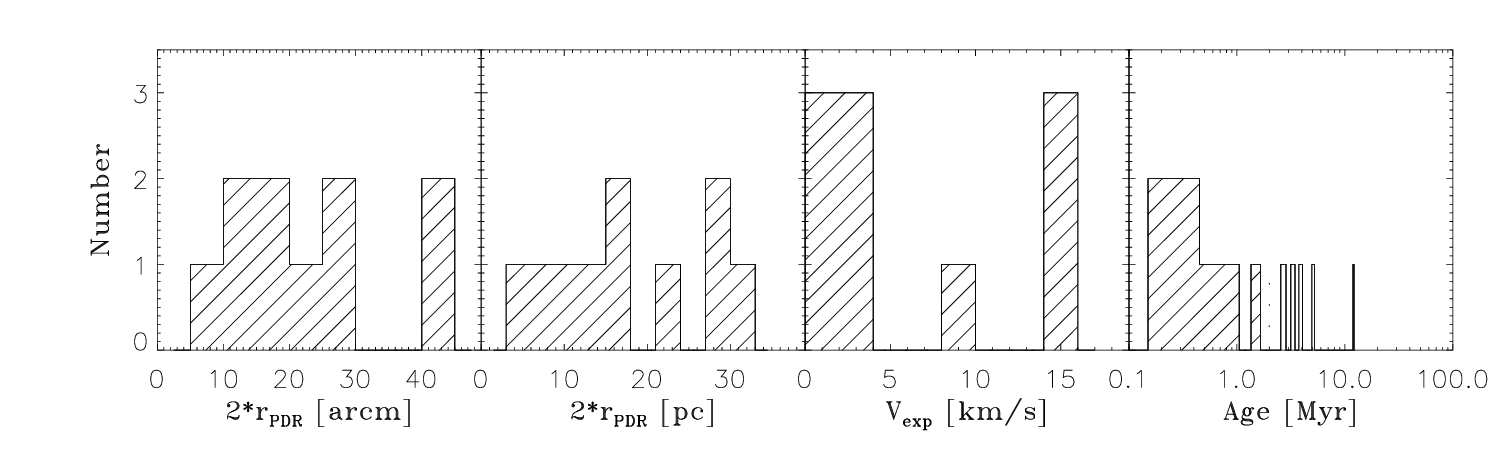}
\caption{Histograms of aperture size, physical size, expansion velocity, and kinematic age. The aperture size values, in units of arcminutes, are taken from the catalog by  \cite{Anderson-2014}, while the other three parameters are derived in this work. The specific values for these parameters are listed in Table \ref{tbl:pars}.}
\label{fig:pars_hist}
\end{figure*}

\subsubsection{Radial profile of flux ratio}
\label{sec:flux_ratio}

For photons with energies higher than the ionization threshold, the absorption cross-section decreases as their energy increases \citep{Osterbrock-2006}. Consequently, the line ratios of [N II]/H$\alpha$, [S II]/H$\alpha$, and [N II]/[S II] are expected to vary with distance from the center of the HII region. Spatially resolved ionization profiles of HII regions provide an ideal laboratory for testing and refining photoionization models.

We therefore explore the flux ratios, including [N II]/H$\alpha$, [S II]/H$\alpha$, and [S II]/[N II], in Figure \ref{fig:flux_ratio_radius}. A similar approach of binning data along the x-axis and applying polynomial fitting, as described in the previous section, is used to fit the flux ratio profile. A default polynomial order of 4 is adopted, except for HII-03, HII-04, and HII-06, where the order is reduced to 2. For comparison, the flux ratio profiles of [N II]/H$\alpha$ and [S II]/H$\alpha$, plotted against the normalized radius ($r/r_{PDR}$), are shown in the bottom-left and bottom-right panels of Figure \ref{fig:profiles}, respectively. In general, the observed trend shows that these ratios increase with distance from the center of the HII regions within the radius of $r_{H\alpha}$.

These trends align with predictions from photoionization models, where the ionization parameter decreases with $r$ \citep{Ferland-2017,Kewley-2019}. Consequently, in the inner regions, ionized metals with higher ionization states dominate, while in the outer regions, lower ionization states become more prevalent \citep{Della_Bruna-2021}. As a result, the [S II]/H$\alpha$ and [N II]/H$\alpha$ ratios increase with $r$  within the region where $r < r_{H\alpha}$. Additionally, due to the different ionization potentials of S (10.3 eV) and N (14.5 eV), the [S II]/[N II] ratio also exhibits spatial variations within the same regions.

\subsubsection{Radial profile of $V_{LSR}$}

The ionized gas is expected to be expelled from its initial location due to the influence of strong radiation fields or stellar winds, with these effects weakening as the radial distance increases. Consequently, the velocity of the ionized gas is predicted to decrease with increasing radial distance.

The radial profile of $V_{LSR}$ are presented in Figure \ref{fig:vlsr_radius}. A similar approach of binning data along the x-axis and applying polynomial fitting, is employed to fit the profile. The default order of the polynomial fit is set to 5, except for HII-03 and HII-04, where it is set to 2, and for HII-06, where it is set to 3.

As each HII region has its own bulk radial velocity, to facilitate comparison, we first subtract the velocity at $r=0.2r_{PDR}$ for each HII regions, and then show these relative velocity profile in the upper-right panel in Figure \ref{fig:profiles}.

In most cases, $V_{LSR}$ at the center is blueshifted relative to that at $r_{PDR}$. This occurs because, under the assumption of ideal spherical symmetry, the far side of the shell, which exhibits redshifted $V_{LSR}$, experiences more significant extinction. As a result, the emission from this region is weaker in the spectrum compared to that from the near side of the shell, which displays blueshifted. Consequently, in spectra with insufficient spectral resolution to distinguish different components along the line of sight, the observed $V_{LSR}$ in the center appears blueshifted relative to that at $r_{PDR}$. However, there are two regions where the velocity profile exhibits a redshift. This result suggests that these two HII regions do not follow a 3D spherical symmetric structure. Instead, they are more likely dominated by the far side of the shell, leading to the observation of only the redshifted component.

\subsection{Escape fraction, expansion velocity and kinematic age}

\subsubsection{Escape fraction}

Under ideal conditions, photons emitted from hot stars are confined within the ionization front, which defines the Str\"omgren radius. The ISM is then divided into an ionized zone and a neutral zone, with the PDR situated between them. However, the ISM surrounding HII regions is non-uniform, leading to irregular shapes that deviate from perfect circles. Additionally, the PDR exhibits patchiness, allowing photons to escape through gaps in its boundaries.

Typically, the fraction of leaked photons can be estimated by comparing the observed photon count, denoted as $Q_{obs}$, with the expected theoretical photon count, denoted as $Q_{exp}$. $Q_{obs}$ is derived from H$\alpha$ imaging flux, while $Q_{exp}$ is calculated based on spatially resolved massive stars or young stellar objects \citep{Della_Bruna-2021}. However, this method faces several challenges: The H$\alpha$ imaging flux includes contributions from the [N II] $\lambda\lambda$6548,6584 emission lines, and the [N II]/H$\alpha$ ratio varies within HII regions, as discussed in Section \ref{sec:flux_ratio}. Additionally, uncertainties arise from background subtraction and reddening corrections. Theoretical calculations of the expected photon count from young stars also introduce further uncertainties.

In this work, we define the escape fraction as the ratio of $F_{out}$  to $F_{tot}$, where $F_{out}$ is the total flux between $r_{PDR}$ and $r_{H\alpha}$, and $F_{tot}$ is  the total flux enclosed within r$_{H\alpha}$, based on the radial profile of  H$\alpha$ flux.
This is calculated as follows:
\begin{equation}
f_{esc} = \frac{F_{out}}{F_{tot}} = \frac{\int_{r_{PDR}}^{r_{H\alpha}}[F(r)-F_c]rdr}{\int_0^{r_{H\alpha}}[F(r)-F_{c}]rdr}
\end{equation}
where $F(r)$  is the H$\alpha$ flux as a function of $r$, and $F_c$ is the continuum flux of H$\alpha$ from the background, defined by the flux at $r_{H\alpha}$. We note that our analysis assumes a uniform filling factor throughout the region.

\subsubsection{Expansion velocity}

In terms of line-of-sight projection, only the radial velocity is observable. Assuming the HII regions undergo spherical expansion, the central region will exhibit the highest radial velocity, while the radial velocity at the PDR corresponds to the bulk motion of the HII region. The difference in velocities between the center of the HII region and the PDR can be interpreted as the expansion velocity ($V_{exp}$). Due to the limited number of data points near the center and to avoid anomalous values at the edges caused by polynomial fitting, we define the velocity at 0.2$r_{PDR}$ as the central velocity of the HII region, denoted as  $V_{cen}$.  The expansion velocity is then calculated as  $V_{exp} = |V_{cen} - V_{PDR}|$. The values of $V_{cen}$, $V_{PDR}$ and $V_{exp}$ are listed in Table \ref{tbl:pars}. $V_{exp}$  ranges from 0.2 $\rm\,km\,s^{-1}$ to 15 $\rm\,km\,s^{-1}$, as shown in Figure \ref{fig:pars_hist}.

\subsubsection{Kinematic age}

It is expected that HII regions expand over time while $V_{exp}$ decreases. As proposed by the model for the evolution of interstellar bubbles \citep{Weaver-1977}, the size of an HII region's bubble scales as  $R(t) \propto L^{1/2} t^{2/5}$, and its expansion velocity scales as $V_{exp}(t)$ $\propto L^{1/2} t^{-3/5}$, where $L$ is the luminosity of the HII region. The kinematic age of the bubble can be estimated using the physical size and expansion velocity of the shell, given by the formula $t = 0.59r_{PDR}/V_{exp}$. The calculated kinematic ages for the HII regions are presented in Table \ref{tbl:pars} and Figure \ref{fig:pars_hist}, ranging from 0.2 to 12 Myr. Notably, the ionization-bounded HII regions exhibit younger ages, with $t$ less than 1 Myr, while the matter-bounded HII regions display older ages. This suggests that matter-bounded HII regions are more evolved compared to the ionization-bounded HII regions in our sample.

\section{Results and discussion}

\subsection{Age dependence of flux profiles}

In the upper-left panel of Figure \ref{fig:profiles}, we present the line profiles of the 10 HII regions, normalized to the flux at $r = 0.2r_{PDR}$. Notably, the two ionization-bounded HII regions exhibit significantly steeper slopes compared to the others. For HII regions with ages between 1-4 Myr, the slopes are relatively shallower and show no clear dependence on age. For the HII region with age of 5 Myr, the slope is even shallower and can be clearly distinguished from the others.  Additionally, the oldest HII region in our sample, with an age of 12 Myr, displays the shallowest slope among all regions studied.

\begin{figure}[ht]
\centering
\includegraphics[width=\linewidth]{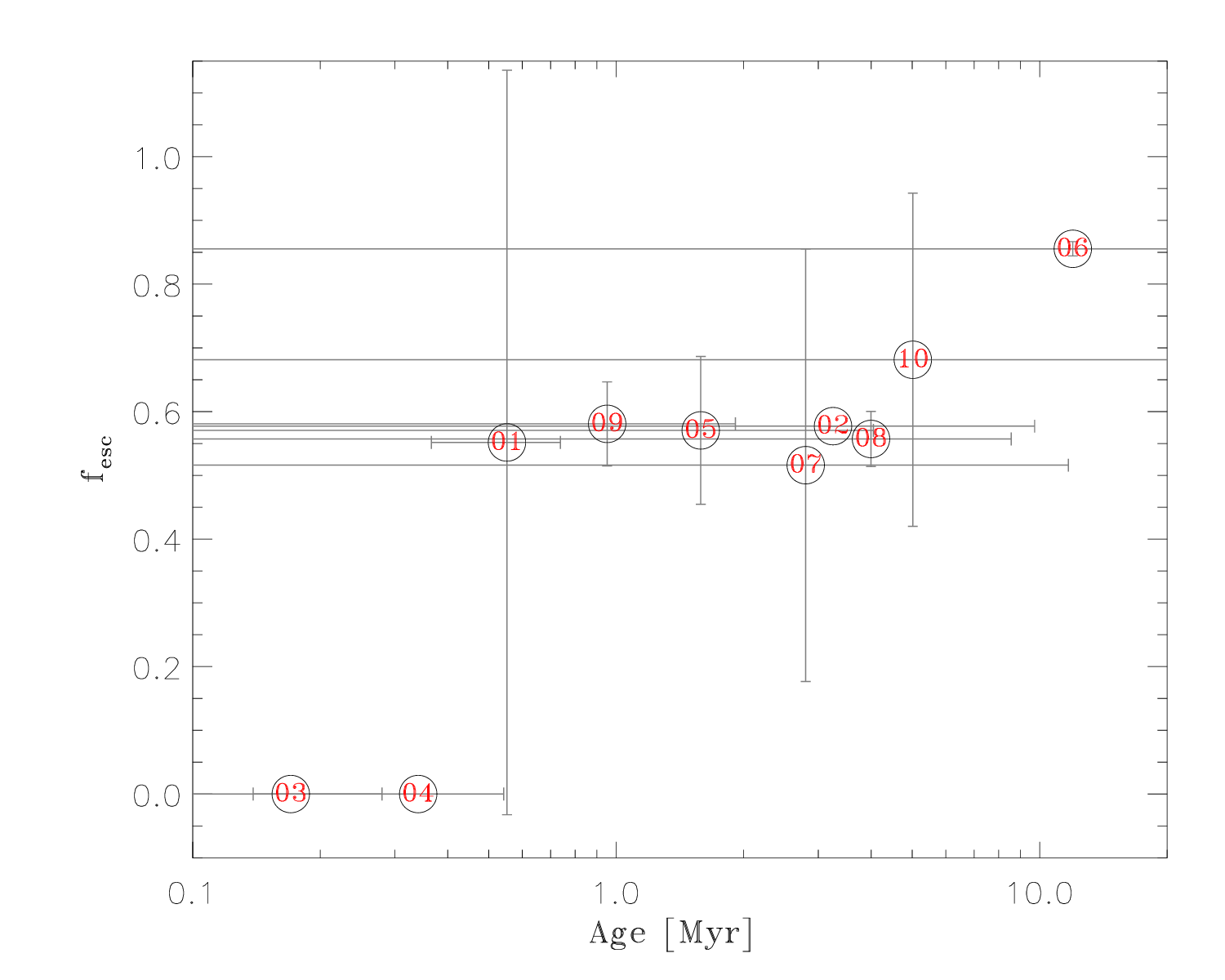}
\caption{Relation between the escape fraction and kinematic age. The ID number for each HII region is enclosed within its corresponding open circle.} 
\label{fig:fesc_t}
\end{figure}

\subsection{Age dependence of $f_{esc}$}

As HII regions expand, photons begin to sweep away the surrounding ISM, leading to an expected increase in the fraction of photons that escape from the HII region and transfer energy into the ISM. The relationship between $f_{esc}$ and age is illustrated in Figure \ref{fig:fesc_t}. It is evident that $f_{esc}$ increases with time, particularly for density-bounded HII regions. In contrast, ionization-bounded HII regions retain photons within their boundaries, resulting in $f_{esc} \sim$ 0. We did not investigate the relationship between $V_{exp}$ and age because $V_{exp}$ is not solely dependent on age but is also influenced by the luminosity of the HII regions, a parameter not explored in this study.

\subsection{Previous works}

A series of studies have analyzed the escape fraction of ionizing photons from HII regions in external galaxies and the Milky Way. \cite{Giammanco-2005} proposed a method to derive $f_{esc}$ by combining the line ratio and the absolute H$\alpha$ luminosity. Applying this method to HII regions in M51, they found that the most plausible values of $f_{esc}$ range between 30\% and 50\%. For some HII regions in M101, $f_{esc}$ can reach up to 60\%. In their study of the optical depth of spatially resolved HII regions in the Magellanic Clouds, \cite{Pellegrini-2012} found that the luminosity-weighted escape fractions from all HII regions are at least 0.42 in the Large Magellanic Cloud (LMC) and 0.40 in the Small Magellanic Cloud (SMC), respectively. These results are consistent with the findings of \cite{Oey-1997}, who suggested that up to 51\% of the ionizing radiation from hot stars escapes the local HII/OB systems in the LMC.

For HII regions in the Milky Way, evidence of photon leakage has also been observed. \cite{Anderson-2015a} surveyed 117 HII regions with multiple RRL velocities, finding that most are clustered near star-forming complexes in the inner Galaxy. They suggested that the additional velocities arise from DIG along the line of sight, likely due to photon leakage from these complexes into the ISM. In a case study, \cite{Anderson-2015b} demonstrated that even for the well-known HII region RCW 120, where the PDR is thick, $\sim$ 25\% of the ionizing photons leak into the surrounding ISM. In another case study, \cite{Luisi-2016} calculated an escape fraction of 15\% for the compact HII region NGC 7538.

There is relatively little research on the radial profiles of emission line fluxes from spatially-resolved HII regions. Here, we highlight two example studies. \cite{Rousseau-Nepton-2018} investigated 4285 HII region candidates in NGC 628 using the Canada–France–Hawaii Telescope (CFHT) imaging spectrograph SITELLE. They quantified the morphologies of HII regions by fitting the H$\alpha$ flux profiles with a pseudo-Voigt function, which combines Gaussian and Lorentzian profiles. \cite{Luisi-2019} analyzed eight Galactic HII regions of varying sizes, morphologies, and luminosities, finding that the hydrogen RRL flux decreases roughly as a power-law with distance from the center. Smaller HII regions exhibit a steeper decrease in intensity, suggesting that giant HII regions play a more significant role in maintaining the ionization of the ISM. However, neither study estimated the ages of these regions nor explored how the profiles evolve with age.

Since leaking photons from HII regions are considered one of the primary sources of ionization for DIG, the escape fraction is a crucial parameter for determining whether photon leakage alone is sufficient or if additional ionization mechanisms are required \citep[e.g.,][]{Oey-2007,Zurita-2002,Della_Bruna-2021,Belfiore-2022}. Our results on how the radial profiles and escape fractions evolve over the lifetime of HII regions provide essential insights into their evolution and can further be applied to model their contribution to the energy budget of the DIG.

\subsection{Limitations}

The current sample of 10 HII regions is relatively small, as it was selected from the dataset completed by the end of 2021. Additionally, while this work focuses on 1D radial profiles, it does not capture important information such as turbulence, outflows, asymmetric distributions, or potential rotation. Therefore, we plan to expand the sample using the latest dataset and analyze the 2D properties in future studies.

As demonstrated, this sample can be considered representative but primarily focuses on relatively small HII regions. This is because giant HII regions are often blended with nearby HII regions and were excluded from this study, as we specifically targeted isolated HII regions. Three HII regions in the sample have physical diameters smaller than 15 pc, resulting in fewer data points within $r_{PDR}$. This introduces larger uncertainties in characterizing the 1D radial profiles of the parameters.

The velocity resolution of 40 $\rm km\,s^{-1}$  from the LAMOST MRS-N dataset is insufficient to distinguish different components along the line of sight. As a result, the expansion velocity calculated in this work is likely underestimated, leading to an overestimation of the kinematic age. Additionally, the Balmer decrement method cannot be applied to this sample for dust extinction correction due to the lack of H$\beta$ emission lines, which affects the accuracy of radial flux profiles to some extent. It would be beneficial to compare these profiles with those derived from RRLs, as RRLs are unaffected by extinction and offer higher velocity resolution.

\section{Summary}

We have constructed a sample of spatially-resolved, isolated Galactic HII regions and analyzed the 1D radial profiles of optical emission flux, flux ratios, and velocities. We have measured various physical parameters of these ten HII regions, including parallax distances, physical sizes, expansion velocities, and escape fractions of photons. The main results are summarized as follows:

\begin{itemize}

\item The distances to these HII regions range from 0.7 to 5.3 kpc. Their physical sizes, as observed in the WISE MIR band, range from 4.8 to 32.6 pc, while the sizes derived from the 1D H$\alpha$ radial flux profiles range from 6.8 to 101.2 pc. This indicates that the optical sizes are approximately 1 to 4 times larger than those observed in the MIR band.

\item The kinematic ages have been estimated based on the physical sizes and expansion velocities of the shells. The two ionization-bounded HII regions are young, with ages of 0.2-0.3 Myr, while the other eight matter-bounded HII regions have ages ranging from 1 to 12 Myr.

\item The emission line flux decreases monotonically with distance from the center of the HII regions. The two ionization-bounded HII regions exhibit the steepest slopes, while the oldest HII region shows a very shallow slope. The slopes of the other regions fall between these extremes.

\item In most cases, the ratios of [N II]/H$\alpha$ and [S II]/H$\alpha$ increase with distance from the center, with the exception of one HII region. Additionally, these ratios vary significantly among different regions, indicating that using a single cutoff value for line ratios is not suitable for distinguishing between HII regions and DIG.

\item The escape fraction $f_{esc}$ of photons from HII regions varies widely, ranging from $\sim$ 0\% to $\sim$ 90\%. For ionization-bounded HII regions, $f_{esc}$ is typically close to 0. In contrast, for matter-bounded HII regions, $f_{esc}$ shows a strong correlation with age: a larger fraction of photons escape as the HII region evolves. This suggests that evolved HII regions older than $\sim$ 1 Myr become the primary energy source for the surrounding DIG.

\end{itemize}

These HII regions can serve as an ideal laboratory for constraining photoionization models through observational comparisons, studying detailed feedback mechanisms on small scales, investigating the expansion mechanisms of HII regions, and characterizing turbulence using optical emission lines.

%
\begin{acknowledgments}

The authors gratefully acknowledge the anonymous referee for the valuable comments and suggestions, which significantly improved the manuscript. W.Z. also thanks Professor Cheng Li and Renbin Yan for their helpful suggestions. This work is supported by the National Natural Science Foundation of China (NSFC) (No. 12090041; 12090044; 12090040) and the National Key R\&D Program of China grant (No. 2021YFA1600401; 2021YFA1600400). This work is also sponsored by the Strategic Priority Research Program of the Chinese Academy of Sciences (No. XDB0550100). The Guoshoujing Telescope (the Large Sky Area Multi-Object Fiber Spectroscopic Telescope LAMOST) is a National Major Scientific Project built by the Chinese Academy of Sciences. Funding for the project has been provided by the National Development and Reform Commission. LAMOST is operated and managed by the National Astronomical Observatories, Chinese Academy of Sciences. 

\end{acknowledgments}


\bibliographystyle{aasjournal}
\bibliography{ms}

\newpage
\appendix
\section{details of various radial profiles} \label{sec:Appendix_profile}

In this appendix, we present the 1D radial distributions of various profiles, including the observed data points and the fitting process using polynomial functions. Specifically: Figure \ref{fig:flux_radius} shows the radial profiles of H$\alpha$, [N II], and [S II] fluxes;
Figure \ref{fig:flux_ratio_radius} displays the line ratios [N II]/H$\alpha$, [S II]/H$\alpha$, and [S II]/[N II]; and
Figure \ref{fig:vlsr_radius} presents the $V_{LSR}$ profiles for H$\alpha$, [N II], and [S II].
The fitting coefficients and errors are summarized in Tables \ref{tbl:coeff1} and \ref{tbl:coeff2}.

\setcounter{figure}{0} 
\renewcommand{\thefigure}{A\arabic{figure}} 

\setcounter{table}{0} 
\renewcommand{\thetable}{A\arabic{table}} 

\begin{figure}[htbp]
\centering
\includegraphics[width=\textwidth]{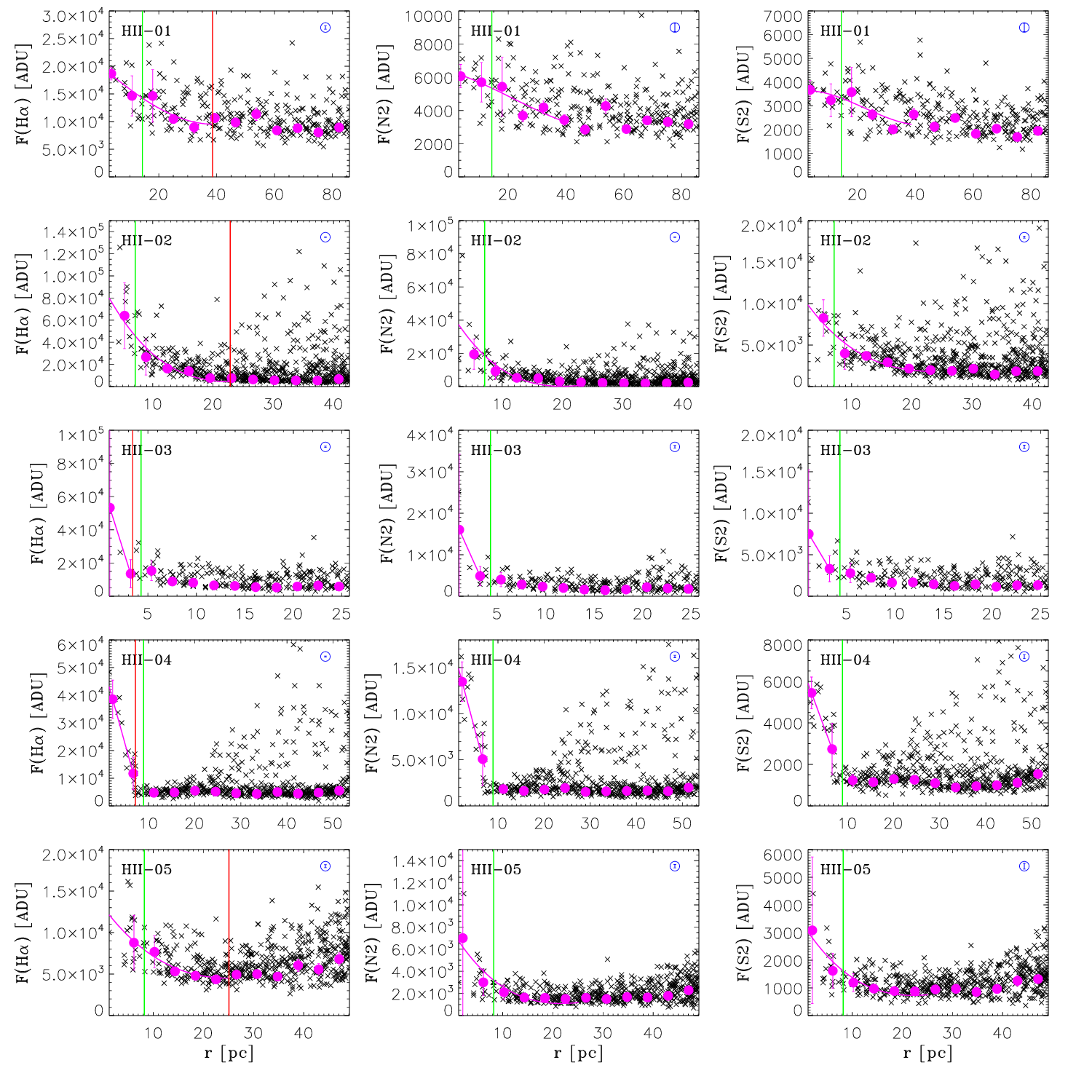}
\caption{1D radial profiles of H$\alpha$, [N II] and [S II] emission line fluxes. Data points located within 6$r_{PDR}$ are represented as black crosses in each panel. The ID number is labeled in the upper-left corner, and typical 1$\sigma$ observational uncertainties are shown in the upper-right corner. For each panel, the data are binned into 10 equally spaced intervals along the x-axis. The median value of each bin, after removing outliers, is represented by magenta-filled circles, with error bars indicating the scatter. The solid magenta line represents the fitting result of the binned data points. The default order of the polynomial fit is set to 4, except for HII-03, HII-04, and HII-06, where it is set to 2. The detailed coefficients and corresponding errors are provided in Table \ref{tbl:coeff1}. The red vertical solid line indicates $r_{H\alpha}$, derived from the 1D radial profile of H$\alpha$ flux, representing the radius at which the H$\alpha$ flux decreases to the background value. The green vertical solid line denotes $r_{PDR}$.}
\label{fig:flux_radius}
\end{figure}

\begin{figure}[htbp]
\centering
\includegraphics[width=\textwidth]{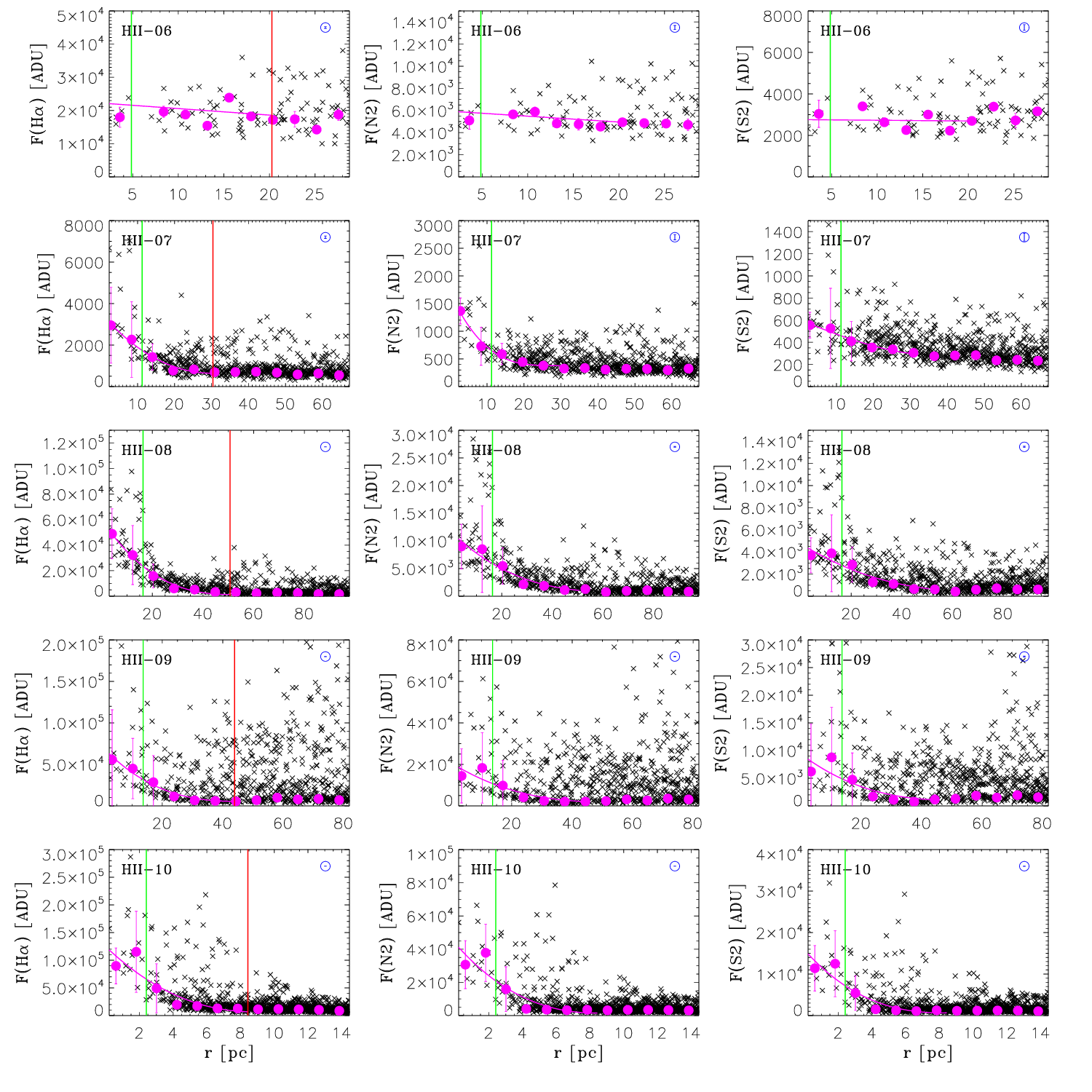}
\continuedcaption{}
\end{figure}

\begin{figure}[htbp]
\centering
\includegraphics[width=\textwidth]{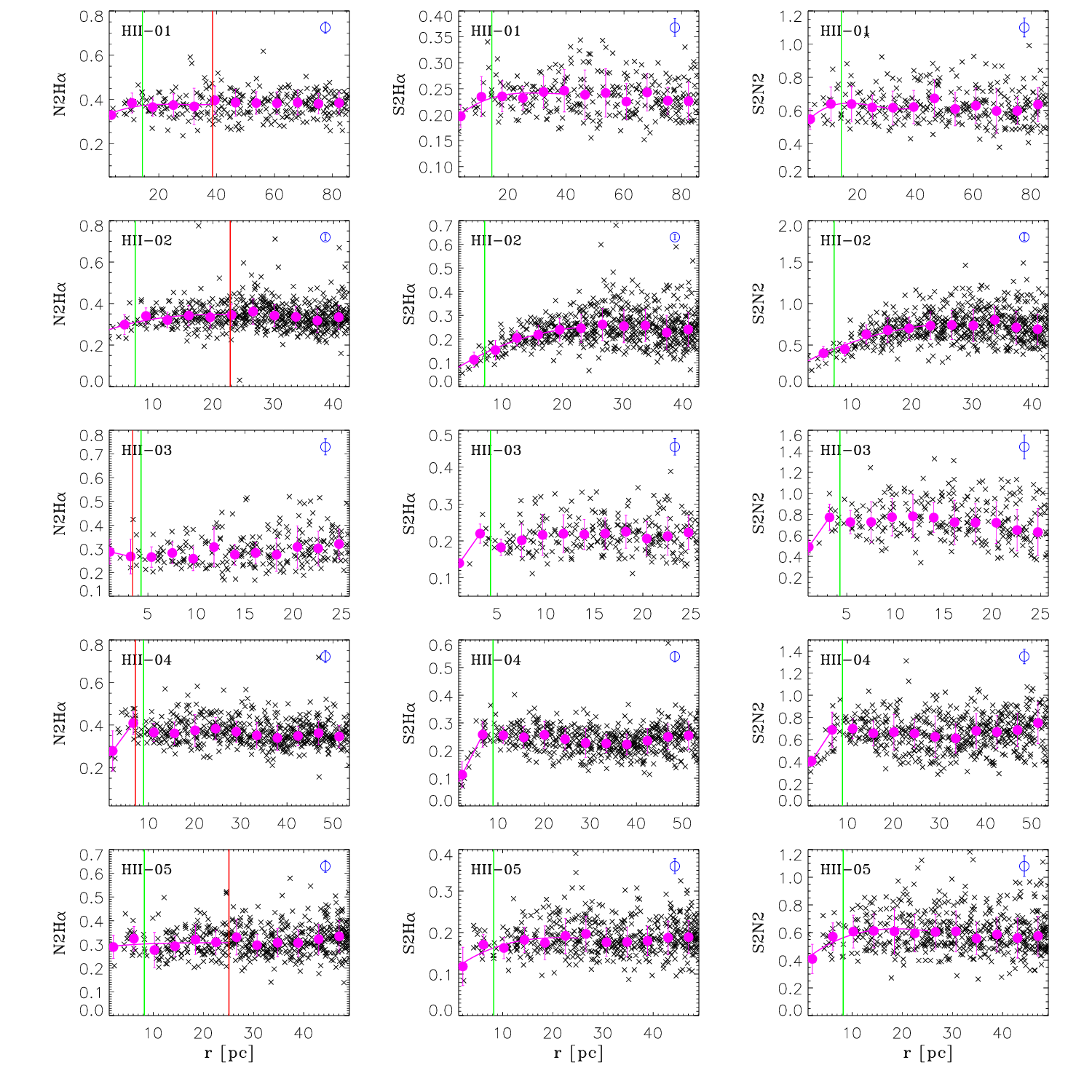}
\caption{1D radial profiles of the line flux ratios [N II]/Ha, [S II]/Ha, and [S II]/[N II]. Data points located within 6$r_{PDR}$ are represented as black crosses in each panel. The ID number is labeled in the upper-left corner, and typical 1$\sigma$ observational uncertainties are shown in the upper-right corner. For each panel, the data are binned into 10 equally spaced intervals along the x-axis. The median value of each bin, after removing outliers, is represented by magenta-filled circles, with error bars indicating the scatter. The solid magenta line represents the fitting result of the binned data points. The default order of the polynomial fit is set to 4, except for HII-03, HII-04, and HII-06, where it is set to 2. The detailed coefficients and corresponding errors are provided in Table \ref{tbl:coeff1}. The red vertical solid line indicates $r_{H\alpha}$, derived from the 1D radial profile of H$\alpha$ flux, representing the radius at which the H$\alpha$ flux decreases to the background value. The green vertical solid line denotes $r_{PDR}$.}
\label{fig:flux_ratio_radius}
\end{figure}

\begin{figure}[htbp]
\centering
\includegraphics[width=\textwidth]{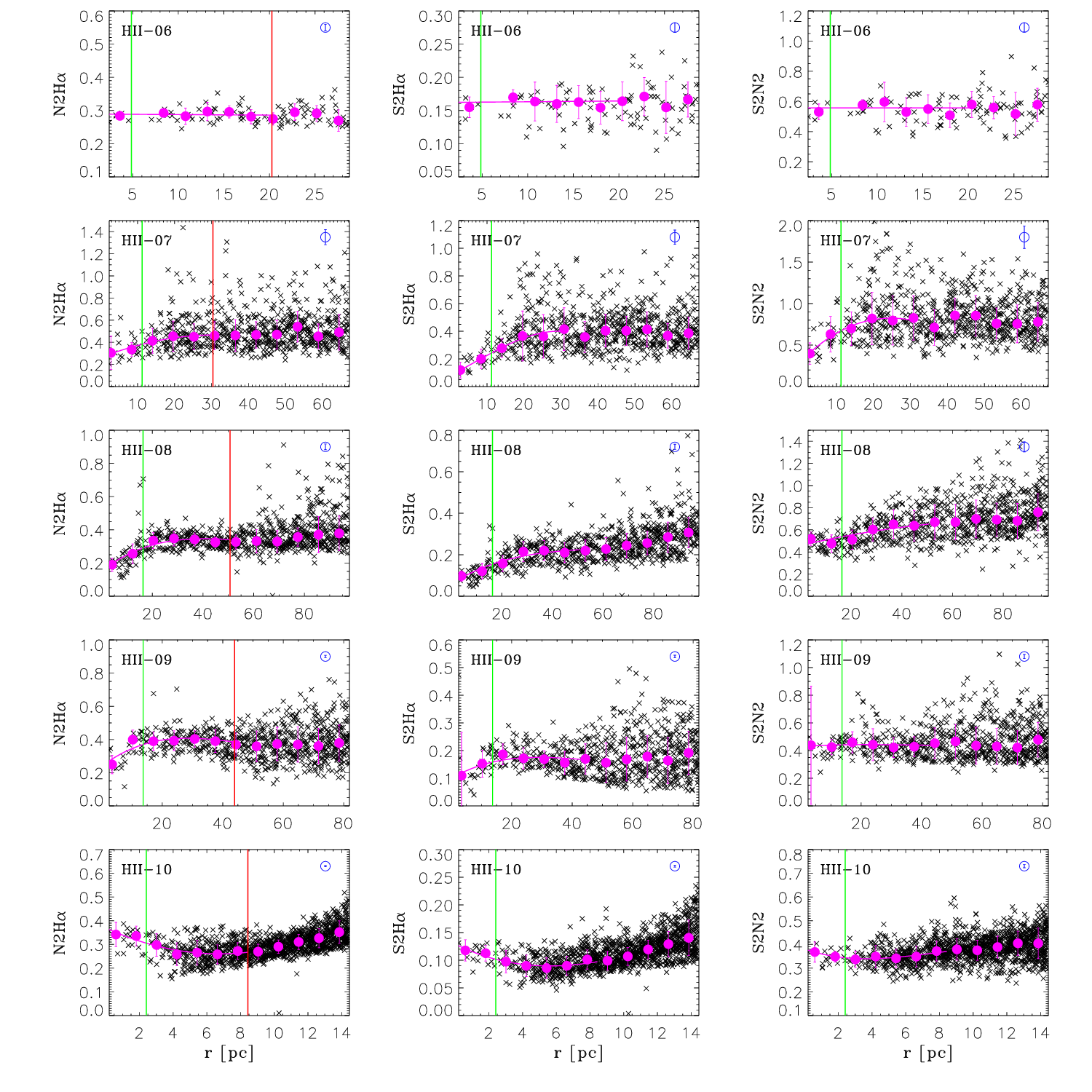}
\continuedcaption{}
\end{figure}

\begin{figure}[htbp]
\centering
\includegraphics[width=\textwidth]{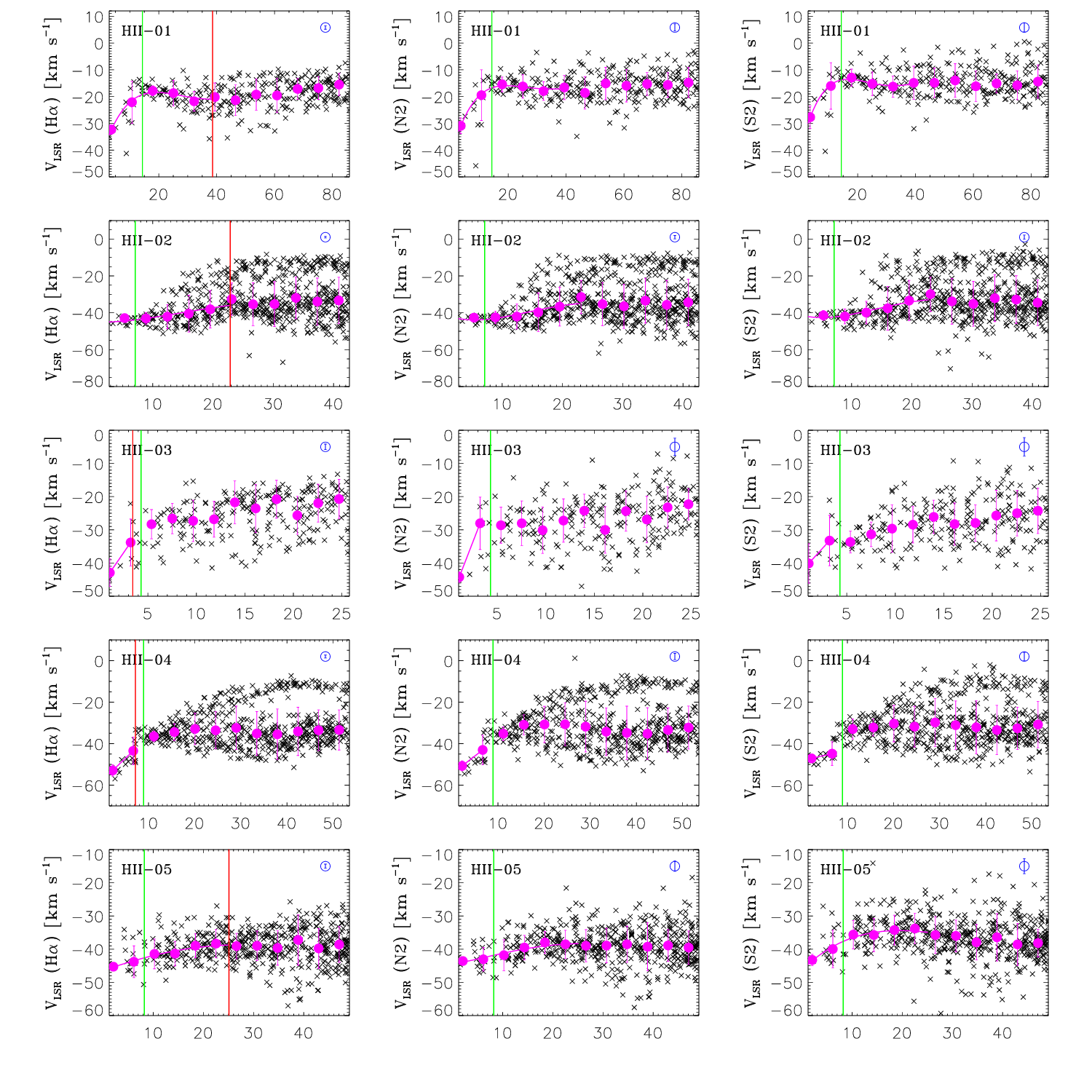}
\caption{1D radial profiles of $V_{LSR}$ for H$\alpha$, [N II] and [S II] emission lines. Data points located within 6$r_{PDR}$ are represented as black crosses in each panel. The ID number is labeled in the upper-left corner, and typical 1$\sigma$ observational uncertainties are shown in the upper-right corner. For each panel, the data are binned into 10 equally spaced intervals along the x-axis. The median value of each bin, after removing outliers, is represented by magenta-filled circles, with error bars indicating the scatter. The solid magenta line represents the fitting result of the binned data points. The default order of the polynomial fit is set to 5, except for HII-03 and HII-04, where it is set to 2, and for HII-06, where it is set to 3. The detailed coefficients and corresponding errors are provided in Table \ref{tbl:coeff2}. The red vertical solid line indicates $r_{H\alpha}$, derived from the 1D profile of H$\alpha$ intensity, representing the radius at which the H$\alpha$ flux decreases to the background value. The green vertical solid line denotes $r_{PDR}$.}
\label{fig:vlsr_radius}
\end{figure}

\begin{figure}[htbp]
\centering
\includegraphics[width=\textwidth]{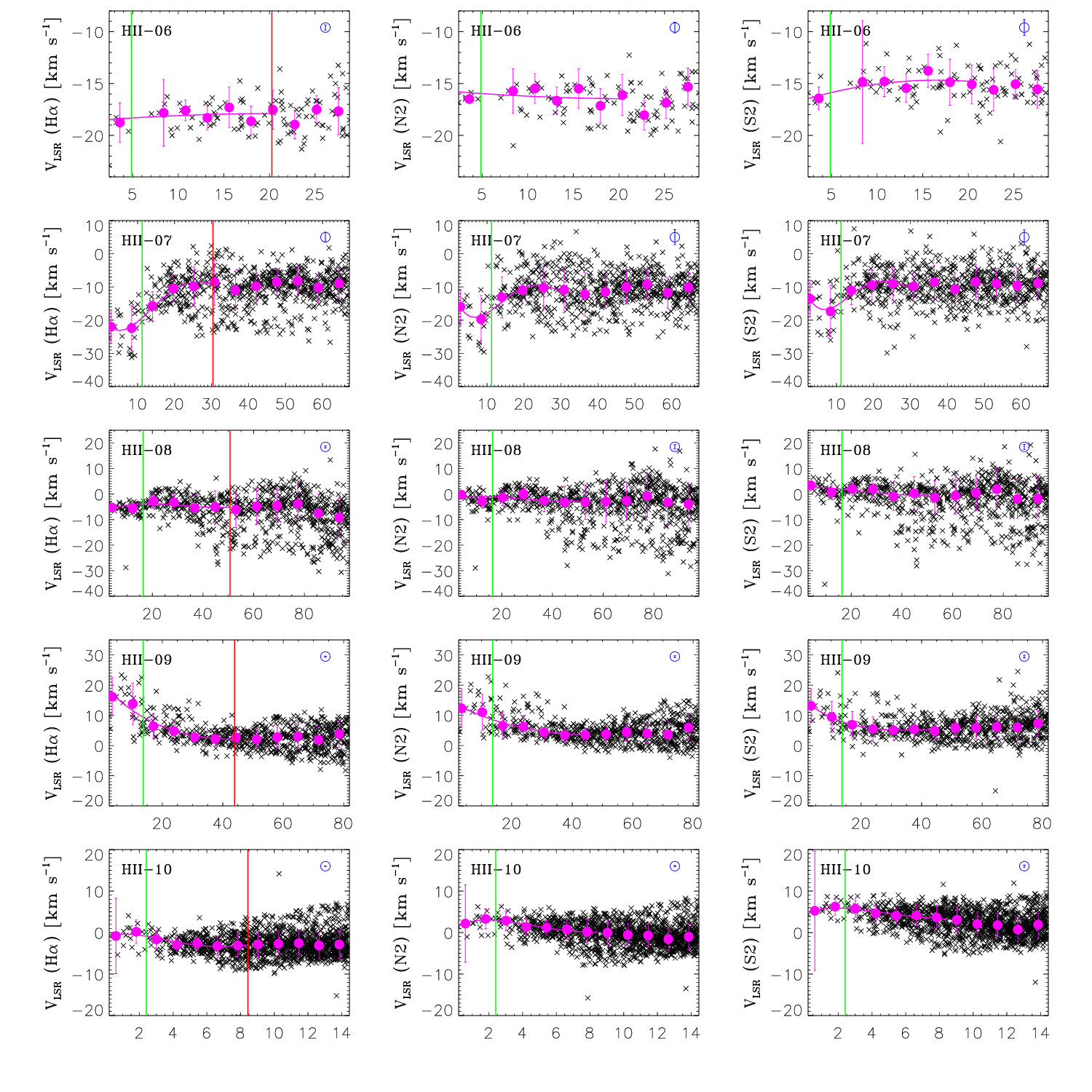}
\continuedcaption{}
\end{figure}

\newpage
\begin{deluxetable*}{cccccccccccccccc}
\tablecaption{Fitting coefficients for the 1D radial profiles of various parameters, including the fluxes of  H$\alpha$ flux, [N II], and [S II], as well as the flux ratios of [N II]/H$\alpha$, [S II]/H$\alpha$, and [S II]/[N II]. The default polynomial order is set to 4, while a linear fitting is applied to HII-03, HII-04, and HII-06. The 1$\sigma$ uncertainties of the coefficients are provided in the line directly below the corresponding coefficient values.  Note: During the fitting process, $r$, the angular distance to the center of the HII region, is given in units of arcdegrees. The radial profile as a function of $r$ (in arcdegrees) has been transformed into a function of $r/r_{PDR}$ (Figure \ref{fig:profiles}) and  projected physical distance (Figures \ref{fig:flux_radius} and \ref{fig:flux_ratio_radius}). \label{tbl:coeff1}}
\tabletypesize{\tiny}
\tablehead{       
  \multicolumn{1}{c}{ID} \vline & \multicolumn{4}{c}{F(H$\alpha$)} \vline& \multicolumn{4}{c}{F(N2)}  \vline& \multicolumn{4}{c}{F(S2)}  \\
\hline      
(1) & (2) & (3) & (4) & (5) & (6) & (7) & (8) & (9) & (10) & (11) &  (12) & (13)
}                 
\startdata
 & a0  &  a1 & a2 & a3 & a0  &  a1 & a2 & a3 & a0  &  a1 & a2 & a3\\
  HII-01 &    20046.5 &   -40694.3 &    16444.9 &    49674.0 &     6356.4 &    -4924.6 &   -12243.6 &    17163.8 &     3618.4 &     1327.7 &   -25936.6 &    35405.0 \\
         & $\pm$346.4 & $\pm$5573.2 & $\pm$23610.9 & $\pm$28298.8 & $\pm$130.3 & $\pm$2096.1 & $\pm$8880.3 & $\pm$10643.5 & $\pm$112.1 & $\pm$1803.1 & $\pm$7639.0 & $\pm$9155.7 \\
  HII-02 &   109686.8 &  -557013.0 &   954611.5 &  -522486.4 &    54097.0 &  -318369.8 &   600336.9 &  -352347.4 &    12811.4 &   -55998.4 &    93003.9 &   -50006.1 \\
         & $\pm$1087.0 & $\pm$8764.8 & $\pm$19992.3 & $\pm$13493.2 & $\pm$385.9 & $\pm$3727.8 & $\pm$9598.5 & $\pm$6991.2 & $\pm$57.4 & $\pm$554.5 & $\pm$1427.8 & $\pm$1040.0 \\
  HII-03 &    73237.3 &  -876319.1 & - & - &    21589.7 &  -245561.1 & - & - &     9598.1 &   -92367.7 & - & - \\
         & $\pm$40372.4 & $\pm$599464.2 & - & - & $\pm$13652.3 & $\pm$201954.0 & - & - & $\pm$5801.0 & $\pm$86729.7 & - & - \\
  HII-04 &    51959.1 &  -243028.9 & - & - &    17667.2 &   -76080.3 & - & - &     6800.0 &   -24532.9 & - & - \\
         & $\pm$5389.2 & $\pm$41083.1 & - & - & $\pm$1765.6 & $\pm$16162.9 & - & - & $\pm$651.4 & $\pm$6581.6 & - & - \\
  HII-05 &    13076.0 &   -75284.9 &   204493.8 &  -160239.3 &     7573.1 &   -67560.0 &   221362.6 &  -218040.2 &     3353.0 &   -26524.5 &    83841.9 &   -78655.7 \\
         & $\pm$106.4 & $\pm$1472.9 & $\pm$5765.7 & $\pm$6678.5 & $\pm$60.3 & $\pm$999.4 & $\pm$4416.4 & $\pm$5520.7 & $\pm$21.4 & $\pm$354.0 & $\pm$1564.4 & $\pm$1955.5 \\
  HII-06 &    22548.3 &   -16128.5 & - & - &     6016.5 &    -4343.9 & - & - &     2767.3 &     -297.9 & - & - \\
         & $\pm$110.7 & $\pm$529.0 & - & - & $\pm$16.0 & $\pm$77.1 & - & - & $\pm$15.4 & $\pm$70.5 & - & - \\
  HII-07 &     3564.6 &    -6657.7 &     4218.5 &     -473.4 &     1700.3 &    -4697.9 &     5608.4 &    -2247.6 &      605.4 &     -451.6 &       61.9 &       73.7 \\
         & $\pm$42.2 & $\pm$307.9 & $\pm$591.9 & $\pm$321.9 & $\pm$17.8 & $\pm$129.9 & $\pm$249.7 & $\pm$135.8 & $\pm$6.4 & $\pm$47.0 & $\pm$90.3 & $\pm$49.1 \\
  HII-08 &    60020.2 &  -133933.7 &    98797.0 &   -23205.1 &    11599.7 &   -19790.1 &    11844.0 &    -2305.2 &     4767.9 &    -6540.4 &     2994.8 &     -371.1 \\
         & $\pm$185.9 & $\pm$750.4 & $\pm$807.3 & $\pm$245.7 & $\pm$77.7 & $\pm$313.4 & $\pm$337.2 & $\pm$102.6 & $\pm$40.8 & $\pm$164.6 & $\pm$177.1 & $\pm$53.9 \\
  HII-09 &    71786.7 &  -238087.4 &   275581.0 &  -101018.2 &    20715.1 &   -56191.5 &    52872.5 &   -15030.8 &     9406.1 &   -25672.3 &    24874.2 &    -7282.7 \\
         & $\pm$324.0 & $\pm$2147.9 & $\pm$3795.5 & $\pm$1897.2 & $\pm$252.6 & $\pm$1675.0 & $\pm$2959.8 & $\pm$1479.5 & $\pm$140.2 & $\pm$929.5 & $\pm$1642.5 & $\pm$821.0 \\
  HII-10 &   124502.0 &  -345974.8 &   288918.4 &   -49159.8 &    43407.8 &  -137310.5 &   136332.5 &   -36704.0 &    15752.5 &   -53705.9 &    60256.9 &   -20525.0 \\
         & $\pm$2487.9 & $\pm$21485.8 & $\pm$49313.1 & $\pm$32017.0 & $\pm$847.7 & $\pm$7321.1 & $\pm$16802.9 & $\pm$10909.5 & $\pm$249.0 & $\pm$2150.3 & $\pm$4935.2 & $\pm$3204.2 \\
\hline
\hline
\multicolumn{1}{c}{ID} \vline & \multicolumn{4}{c}{N2Ha} \vline& \multicolumn{4}{c}{S2Ha}  \vline& \multicolumn{4}{c}{S2N2}  \\
\hline
 & a0  &  a1 & a2 & a3 & a0  &  a1 & a2 & a3 & a0  &  a1 & a2 & a3\\
  HII-01 &      0.315 &      0.649 &     -2.169 &      2.343 &      0.184 &      0.494 &     -1.363 &      1.209 &      0.484 &      2.046 &     -8.145 &      9.545 \\
         & $\pm$0.004 & $\pm$0.071 & $\pm$0.299 & $\pm$0.358 & $\pm$0.002 & $\pm$0.036 & $\pm$0.154 & $\pm$0.185 & $\pm$0.003 & $\pm$0.043 & $\pm$0.182 & $\pm$0.218 \\
  HII-02 &      0.245 &      0.557 &     -0.945 &      0.478 &      0.035 &      0.841 &     -0.968 &      0.301 &      0.204 &      1.930 &     -1.867 &      0.335 \\
         & $\pm$0.001 & $\pm$0.013 & $\pm$0.033 & $\pm$0.024 & $\pm$0.001 & $\pm$0.008 & $\pm$0.022 & $\pm$0.016 & $\pm$0.003 & $\pm$0.033 & $\pm$0.085 & $\pm$0.062 \\
  HII-03 &      0.297 &     -0.431 & - & - &      0.100 &      1.766 & - & - &      0.349 &      6.198 & - & - \\
         & $\pm$0.043 & $\pm$0.986 & - & - & $\pm$0.007 & $\pm$0.288 & - & - & $\pm$0.064 & $\pm$1.610 & - & - \\
  HII-04 &      0.213 &      1.185 & - & - &      0.040 &      1.315 & - & - &      0.263 &      2.563 & - & - \\
         & $\pm$0.072 & $\pm$0.488 & - & - & $\pm$0.037 & $\pm$0.294 & - & - & $\pm$0.053 & $\pm$0.753 & - & - \\
  HII-05 &      0.289 &      0.217 &     -0.868 &      1.193 &      0.109 &      0.845 &     -2.889 &      3.046 &      0.383 &      2.877 &    -10.497 &     11.075 \\
         & $\pm$0.002 & $\pm$0.029 & $\pm$0.130 & $\pm$0.162 & $\pm$0.001 & $\pm$0.016 & $\pm$0.073 & $\pm$0.091 & $\pm$0.003 & $\pm$0.044 & $\pm$0.195 & $\pm$0.244 \\
  HII-06 &      0.289 &     -0.012 & - & - &      0.162 &      0.009 & - & - &      0.557 &      0.002 & - & - \\
         & $\pm$0.001 & $\pm$0.006 & - & - & $\pm$0.001 & $\pm$0.008 & - & - & $\pm$0.004 & $\pm$0.026 & - & - \\
  HII-07 &      0.262 &      0.415 &     -0.192 &     -0.019 &      0.079 &      0.440 &      0.203 &     -0.330 &      0.289 &      1.475 &     -1.108 &      0.123 \\
         & $\pm$0.005 & $\pm$0.033 & $\pm$0.064 & $\pm$0.035 & $\pm$0.005 & $\pm$0.038 & $\pm$0.074 & $\pm$0.040 & $\pm$0.009 & $\pm$0.067 & $\pm$0.130 & $\pm$0.070 \\
  HII-08 &      0.151 &      0.544 &     -0.484 &      0.135 &      0.058 &      0.355 &     -0.270 &      0.076 &      0.457 &      0.252 &     -0.084 &      0.012 \\
         & $\pm$0.002 & $\pm$0.007 & $\pm$0.007 & $\pm$0.002 & $\pm$0.001 & $\pm$0.005 & $\pm$0.005 & $\pm$0.002 & $\pm$0.004 & $\pm$0.014 & $\pm$0.015 & $\pm$0.005 \\
  HII-09 &      0.239 &      0.868 &     -1.372 &      0.617 &      0.099 &      0.392 &     -0.632 &      0.302 &      0.428 &      0.095 &     -0.190 &      0.106 \\
         & $\pm$0.003 & $\pm$0.017 & $\pm$0.030 & $\pm$0.015 & $\pm$0.001 & $\pm$0.008 & $\pm$0.015 & $\pm$0.007 & $\pm$0.002 & $\pm$0.014 & $\pm$0.024 & $\pm$0.012 \\
  HII-10 &      0.369 &     -0.386 &      0.344 &     -0.005 &      0.129 &     -0.180 &      0.231 &     -0.058 &      0.381 &     -0.296 &      0.632 &     -0.331 \\
         & $\pm$0.002 & $\pm$0.013 & $\pm$0.030 & $\pm$0.020 & $\pm$0.001 & $\pm$0.005 & $\pm$0.012 & $\pm$0.008 & $\pm$0.001 & $\pm$0.008 & $\pm$0.018 & $\pm$0.011 \\
\hline
\enddata 
\end{deluxetable*}  

\begin{deluxetable*}{cccccccccccccccccccccc}
\tablecaption{Fitting coefficients for the 1D radial profiles of V$_{LSR}$ for the emission lines of H$\alpha$, [N II], and [S II]. The default polynomial order is set to 5. A linear fitting is applied to HII-03 and HII-04, while a polynomial of order 3 is used for HII-06. The 1$\sigma$ uncertainties of the coefficients are provided in the line directly below the corresponding coefficient values. Note: During the fitting process, $r$, the angular distance to the center of the HII region, is given in units of arcdegrees. The radial profile as a function of $r$ (in arcdegrees) has been transformed into a function of $r/r_{PDR}$ (Figure \ref{fig:profiles}) and  projected physical distance (Figure \ref{fig:vlsr_radius}).
\label{tbl:coeff2}}
\tabletypesize{\tiny}
\tablehead{       
\multicolumn{1}{c}{ID} \vline & \multicolumn{5}{c}{V$_{LSR}$ (H$\alpha$)} \vline& \multicolumn{5}{c}{V$_{LSR}$ (N2)}  \vline& \multicolumn{5}{c}{V$_{LSR}$ (S2)}  \\
\hline      
(1) & (2) & (3) & (4) & (5) & (6) & (7) & (8) & (9) & (10) & (11) &  (12) & (13) & (14) & (15) & (16)
}                 
\startdata
\hline
 & a0  &  a1 & a2 & a3 & a4 &  a0  &  a1 & a2 & a3 & a4 & a0  &  a1 & a2 & a3 & a4\\
  HII-01 &      -42.9 &      315.9 &    -1421.6 &     2596.9 &    -1686.6 &      -42.9 &      366.9 &    -1778.0 &     3623.8 &    -2673.6 &      -40.7 &      404.3 &    -2087.1 &     4352.1 &    -3173.0 \\
         & $\pm$0.5 & $\pm$13.1 & $\pm$95.9 & $\pm$261.4 & $\pm$236.5 & $\pm$0.3 & $\pm$9.0 & $\pm$65.9 & $\pm$179.6 & $\pm$162.4 & $\pm$0.3 & $\pm$6.8 & $\pm$49.5 & $\pm$134.8 & $\pm$121.9 \\
  HII-02 &      -45.0 &       -5.8 &      113.5 &     -160.6 &       63.4 &      -40.4 &      -68.1 &      395.5 &     -633.9 &      320.0 &      -39.5 &      -63.9 &      392.1 &     -631.2 &      314.4 \\
         & $\pm$0.2 & $\pm$3.6 & $\pm$16.1 & $\pm$26.8 & $\pm$14.7 & $\pm$0.3 & $\pm$3.9 & $\pm$17.5 & $\pm$29.0 & $\pm$15.9 & $\pm$0.3 & $\pm$3.9 & $\pm$17.6 & $\pm$29.2 & $\pm$16.1 \\
  HII-03 &      -47.4 &      200.4 & - & - & - &      -52.3 &      357.0 & - & - & - &      -43.6 &      151.7 & - & - & - \\
         & $\pm$3.6 & $\pm$90.2 & - & - & - & $\pm$2.3 & $\pm$88.6 & - & - & - & $\pm$4.8 & $\pm$106.0 & - & - & - \\
  HII-04 &      -57.6 &       85.6 & - & - & - &      -54.6 &       69.7 & - & - & - &      -48.3 &       20.8 & - & - & - \\
         & $\pm$2.7 & $\pm$32.6 & - & - & - & $\pm$2.5 & $\pm$33.9 & - & - & - & $\pm$2.1 & $\pm$27.7 & - & - & - \\
  HII-05 &      -46.4 &       48.2 &      -36.1 &     -227.4 &      336.4 &      -44.5 &       25.7 &      103.1 &     -596.6 &      659.7 &      -46.3 &      138.6 &     -517.2 &      646.0 &     -202.0 \\
         & $\pm$0.1 & $\pm$3.4 & $\pm$25.9 & $\pm$73.9 & $\pm$69.7 & $\pm$0.1 & $\pm$2.4 & $\pm$18.6 & $\pm$53.0 & $\pm$50.0 & $\pm$0.1 & $\pm$3.2 & $\pm$24.6 & $\pm$70.0 & $\pm$66.1 \\
  HII-06 &      -18.6 &        5.7 &      -11.1 & - & - &      -15.6 &       -6.1 &       10.9 & - & - &      -17.1 &       24.1 &      -59.8 & - & - \\
         & $\pm$0.1 & $\pm$1.1 & $\pm$2.7 & - & - & $\pm$0.1 & $\pm$1.6 & $\pm$4.0 & - & - & $\pm$0.1 & $\pm$0.9 & $\pm$2.3 & - & - \\
  HII-07 &      -18.0 &      -71.2 &      297.5 &     -338.3 &      120.2 &       -9.1 &     -112.5 &      399.7 &     -464.2 &      174.0 &       -6.4 &     -117.9 &      423.5 &     -510.2 &      200.4 \\
         & $\pm$0.5 & $\pm$5.8 & $\pm$19.1 & $\pm$23.6 & $\pm$9.7 & $\pm$0.5 & $\pm$6.4 & $\pm$21.3 & $\pm$26.3 & $\pm$10.8 & $\pm$0.5 & $\pm$5.9 & $\pm$19.5 & $\pm$24.1 & $\pm$9.9 \\
  HII-08 &       -7.6 &       23.4 &      -46.6 &       33.8 &       -8.2 &       -1.5 &        6.2 &      -19.5 &       16.0 &       -4.0 &        2.7 &        2.1 &      -17.1 &       16.3 &       -4.4 \\
         & $\pm$0.2 & $\pm$1.0 & $\pm$1.9 & $\pm$1.3 & $\pm$0.3 & $\pm$0.2 & $\pm$1.1 & $\pm$2.1 & $\pm$1.4 & $\pm$0.3 & $\pm$0.2 & $\pm$1.2 & $\pm$2.2 & $\pm$1.5 & $\pm$0.4 \\
  HII-09 &       20.5 &      -66.2 &       80.6 &      -37.3 &        5.1 &       14.6 &      -34.4 &       34.8 &      -13.4 &        2.3 &       16.1 &      -56.4 &      103.6 &      -81.3 &       23.9 \\
         & $\pm$0.2 & $\pm$1.7 & $\pm$5.1 & $\pm$5.9 & $\pm$2.2 & $\pm$0.1 & $\pm$1.2 & $\pm$3.7 & $\pm$4.2 & $\pm$1.6 & $\pm$0.0 & $\pm$0.4 & $\pm$1.3 & $\pm$1.5 & $\pm$0.6 \\
  HII-10 &       -1.1 &       14.4 &      -94.6 &      150.3 &      -72.1 &        1.4 &       23.2 &     -101.5 &      135.6 &      -59.9 &        4.7 &       17.0 &      -75.9 &      102.0 &      -46.8 \\
         & $\pm$0.1 & $\pm$1.4 & $\pm$5.7 & $\pm$8.4 & $\pm$4.1 & $\pm$0.1 & $\pm$1.0 & $\pm$3.8 & $\pm$5.6 & $\pm$2.8 & $\pm$0.1 & $\pm$1.1 & $\pm$4.2 & $\pm$6.3 & $\pm$3.1 \\
\hline
\enddata 
\end{deluxetable*}  

\label{lastpage}
\end{document}